\documentclass[english,notitlepage,nofootinbib]{revtex4-1}
\usepackage[T1]{fontenc}
\usepackage[latin9]{inputenc}
\usepackage{color}
\usepackage{amsmath}
\usepackage{amssymb}
\usepackage{graphicx}

\makeatletter

\providecommand{\tabularnewline}{\\}

\usepackage{xcolor}

\makeatother

\usepackage{babel}
\begin{document}

%%%%%%%%%%%%%%%%%%%%%%%%%%%%%%%%%%%%%%%%%%%%%%%%%%%%%%%%%%%%%%%%%%%%%%
\title{Neutrino mixing in SO(10) GUTs with non-abelian flavor symmetry in the hidden sector}
%%%%%%%%%%%%%%%%%%%%%%%%%%%%%%%%%%%%%%%%%%%%%%%%%%%%%%%%%%%%%%%%%%

\author{Alexei Yu. Smirnov and Xun-Jie Xu}

\affiliation{\textcolor{black}{Max-Planck-Institut f\"ur Kernphysik, Postfach
103980, D-69029 Heidelberg, Germany}.}

\date{\today}
\begin{abstract}
The relation between the mixing matrices of leptons and quarks:
$U_{\rm{PMNS}} \approx V_{\rm{CKM}}^\dagger U_0$, where $U_0$ is a matrix of
special forms (e.g. BM, TBM), can be a clue for understanding
the lepton mixing and neutrino masses.
It may imply the Grand unification and existence of a hidden  sector with certain symmetry which generates
$U_0$ and leads to the smallness of neutrino masses. We apply the residual symmetry
approach to obtain $U_0$.
The residual symmetries of both the visible and hidden sectors are
$\mathbb{Z}_{2} \times \mathbb{Z}_{2}$.
Their embedding in a unified flavor group is considered.
We find that there are only several possible structures of $U_0$,
including the BM mixing and matrices with elements determined by the golden ratio.
Realization of the BM scenario based on the  $SO(10)$ GUT with  the $S_4$
flavor group is presented.  Generic features of
this scenario are discussed, in particular, the prediction of CP phase
$144^{\circ}\lesssim\delta_{\rm CP}\lesssim 210^{\circ}$ in the minimal version.

\end{abstract}
\maketitle

%%{\color{lightgray}

\section{Introduction}
%%%%%%%%%%%%%%%%%%%%%%%%%%%%%%%%%%%%%%%%%%%%%%%%%%%%%%%%%%%%%%%%%%%%%%%%%%%%

There is an  appealing approximate relation  between the mixing matrices of leptons,
$U_{\rm PMNS}$,   and quarks, $V_{\rm CKM}$ \cite{Giunti:2002ye,Raidal:2004iw,Minakata:2004xt,
Frampton:2004vw,Ferrandis:2004vp}:
\begin{equation}
U_{\rm PMNS} \simeq V_{\rm CKM}^\dagger U_0,
\label{eq:gut-45}
\end{equation}
where $U_0$ is close to the  bi-maximal (BM,
$U_{\rm{BM}}$) \cite{Vissani:1997pa,Barger:1998ta},
%,Mohapatra:1998ka,Nomura:1998gm,
%Frampton:2004ud,Altarelli:2009gn,He:2012yt},
or tri-bi-maximal
(TBM, $U_{\rm{TBM}}$) \cite{Harrison:2002er,Harrison:2002kp}
%Xing:2002sw, Harrison:2002et,Harrison:2003aw
mixing matrices.
In particular, Eq.~(\ref{eq:gut-45}) gives a relation between
the Cabibbo angle $\theta_C$
and the leptonic mixing angles $\theta_{13}$ and $\theta_{23}$:
\begin{equation}
\sin^2\theta_{13} \simeq
\sin^2\theta_{23} \sin^2\theta_C.
%\frac{\sin^2\theta_C}{2}.
%\left[ .... \right].
\label{eq:13-cabi}
\end{equation}
The relation  (\ref{eq:gut-45}) is in a good agreement with
available experimental results \cite{Esteban:2016qun}
and has been widely studied in the literature
\cite{Cheung:2005gq,Xing:2005ur,Datta:2005ci,Antusch:2005kw,
Harada:2005km,
Antusch:2005ca,Li:2005ir,Schmidt:2006rb,Hochmuth:2006xn,
Plentinger:2007px,
Zhang:2012pv,Zhang:2012zh,
Harada:2013aja,Ke:2014hxa,Sharma:2015dqi,Sharma:2017wcv}.
If not accidental, it can be the clue for understanding
peculiar features of the lepton mixing,  and eventually,
the origins of neutrino masses.
The following logical steps lead to a rather restricted scenario. \\

1. The relation  (\ref{eq:gut-45}) implies that leptons ``know'' about quarks.
It should be  a kind of quark-lepton unification, probably
the Grand Unification at high energy scale.  Indeed, the Grand Unification can
ensure similarity of the Dirac mass matrices
of quarks and leptons: $m_D^\nu \sim m_D^{\rm up}$ and $ m_D^l \sim m_D^{\rm down}$.
This leads to  appearance of mixing $ \sim V_{\rm{CKM}}$ in the lepton sector.\footnote{
Alternatively,  the relations between the mass matrices can be obtained as a
consequence of common flavor symmetry in both sectors.}

2. At the same time the difference between the quark and lepton mixing implies existence of some
new physics responsible for generation of matrix $U_0$.
The structure of this matrix indicates certain underlying symmetry which is difficult
to extend to the quark sector.

3. It is natural to assume that the same new physics is responsible for $U_0$ and smallness of neutrino masses.
In the Grand Unification framework the simplest way to get small neutrino masses is to
invoke the high mass scale type I seesaw mechanism
\cite{Minkowski:1977sc,yanagida1979proceedings,GellMann:1980vs,
glashow1979future,mohapatra1980neutrino}:
\begin{equation}
m_\nu = - m_D^{\nu} \frac{1}{M_R} m_D^{\nu T},
\label{eq:}
\end{equation}
where  $M_R$ is the mass matrix of right-handed (RH) neutrino components.
%%that leads to appearance of $V_{\rm{CKM}}$ in both sectors.
Thus, in the seesaw mechanism,  ${M_R}$  with specific properties
could be responsible for generation of  $U_0$.
%%Then aperance of
%%$U_0$ can be associated to additional structure
%%in lepton sector - Majorana mass matrix of RH
%%neutrinos $M_R$.
%%

4. The latter, however, implies very strong (quadratic) hierarchy
of masses of the neutrinos and enormous fine tuning which is very
difficult (if possible) to justify in the usual seesaw mechanism.
%%It is difficult to realize required properties of $M_R$
%%(which lead to $U_{\rm{BM}}$ or $U_{\rm{TBM}}$)
%%in the minimal version.
One way to solve this problem is to introduce the double seesaw
mechanism in which the RH neutrinos themselves acquire masses via the seesaw mechanism \cite{Mohapatra:1986bd}.
%,Mohapatra:1986aw
This opens up a possibility to cancel the strong hierarchy
as a result of certain  symmetry \cite{Lindner:2005pk,Hagedorn:2008bc}.
%%The double seesaw mechanism requires introduction of new Majorana fermions $S$.
%%If the mass terms, which mix $S$ with RH neutrinos, are at the GUT scale (at least the largest mass):
%%$M_{SN} \sim M_{\rm{GUT}} \sim 10^{16}$ GeV, then correct masses of light neutrinos
%%are generated when  the masses of singlets $S$ are at the string-Planck scale $M_{S} \sim M_{\rm{string}}
%%\sim 10^{18}$ GeV.
%%
Furthermore, the structure of mass matrix of $S$, governed by certain
symmetry can eventually lead to the required mixing $U_0$.
%%So, the double seesaw allows  to implement two symmetries: one leads to cancellation
%%of strong hierarchies in the Dirac mass matrices; another one is realized in the new sector of
%%$S$-fields generates $U_{0}$.

These general arguments can be realized in the following scenario. \\ %[[move to sect II? ]]

\begin{itemize}

\item

There is the Grand Unification based on $SO(10)$ gauge symmetry group \cite{Georgi:1974my,Fritzsch:1974nn} with fermions
in 16-plet representations which include also the RH neutrinos. This ensures
similarity of the Dirac  mass matrices of the quarks and leptons and also
the coincidence of scales $M_{SN} \sim M_{\rm{GUT}}$.

\item

A hidden sector exists which consists of singlet fermions and bosons of $SO(10)$.  This sector couples with the
visible one via the RH neutrino portal. The fermions $S$ that participate in the double seesaw mechanism belong to this
sector. Symmetries of the hidden sector lead to the matrix $U_0$ with required properties.

\item

Information about mixing in the hidden sector should be communicated to the visible sector. The minimal possibility
is to fix basis of states in all sectors (visible, portal, hidden), and this can be done by introducing the basis fixing
symmetry \cite{Ludl:2015tha,Chu:2016lkb,Bajc:2016eiw}. In the case of three generations the simplest possibility is the
$\mathbb{Z}_{2}\times\mathbb{Z}_{2}$ symmetry \cite{Ludl:2015tha}. In turn, such a symmetry can be a part of intrinsic
symmetry of theory which is always present. The basis fixing symmetry ensures that all mass matrices of the visible
sector and portal are diagonal. This symmetry is spontaneously broken in the hidden sector by interactions with flavons
leading to another unbroken (intrinsic) $\mathbb{Z}_{2}\times\mathbb{Z}_{2}$ and generating
$U_0$. No such a structure exists in the quark sector.

%%flavor symmetry+fit {[}1306.4468{]},...

\item

Additional  physics should be introduced to generate the CKM mixing.

\item

In the visible sector $\mathbb{Z}_{2}\times\mathbb{Z}_{2}$ is broken by another
mechanism leading to the CKM mixing.
The double seesaw mechanism allows to  disentangle generation of CKM and  $U_0$ mixings.
Furthermore,  introducing the hidden sector allows one to construct
economical renormalizable theory  with flavor symmetry.

\end{itemize}

In general, $\mathbb{Z}_{2}\times\mathbb{Z}_{2}$ can lead to large mixing
in $U_0$, but it does not produce specific structures such as BM or TBM.
To this end, non-abelian symmetry should be introduced in the hidden sector. In other words,
the basis symmetry $\mathbb{Z}_{2}\times\mathbb{Z}_{2}$ should be promoted to non-abelian symmetry.
In this paper we study such a possibility.
We focus on the symmetry issues:
the interplay between the gauge $SO(10)$ and discrete
flavor symmetries.
%independent of the model construction.
We also consider generation of $U_l\simeq V_{\rm{CKM}}$.
 %the UV completion, hierarchy problem, etc. is beyond the scope of this paper. )

%%In other terms: ... hidden sector portal...
%%Generation of large lepton mixing in the hidden sector
%%allows  renormalizable theory

The paper is organized as follows. In sect. \ref{sec:Framework} we describe the scenario in details.
In sect. \ref{sec:symmetry} we  study the possibility to generate the matrix $U_0$
using the residual symmetry approach
applied to the visible and hidden sectors.
We find all possible structures of $U_0$. In sect. \ref{sec:BM} we present
realization the residual symmetry mechanism which generates $U_0 = U_{\rm{BM}}$
 and is based on $S_4$ symmetry group. We explore a possibility to generate the CKM mixing
 and study the predictions for the PMNS mixing in sect. \ref{sec:CKM}.
 % \ref{sec:Prediction}.
%Generic aspects features of the model are discussed in sect VI.
Conclusions are given in sect. \ref{sec:Conclusion}.

%\xxj \xj{references are not enough, to be added...}
%e.g. S4:
%Altarelli:2009gn
%Shimizu:2015tta,Zhang:2015vle,Li:2014eia,Ding:2013eca
%%}

\section{Framework\label{sec:Framework}}
%%%%%%%%%%%%%%%%%%%%%%%%%%%%%%%%%%%%%%%%%%%%%%%%%%%%%%%%%%%%%%%%%%%%%%%%%%

Let us describe the main elements of the framework.
%%In this section, we briefly review the framework  to accommodate
%%large neutrino mixing in the hidden sector of $SO(10)$ grand unification,
%%which has been previously studied in Refs.~\cite{Ludl:2015tha,Chu:2016lkb,Bajc:2016eiw}.

{\it 1. Visible, portal and hidden sectors.}
The visible sector includes  3 families of fermions accommodated in three 16-plets of $SO(10)$:
$(\psi_{1},\thinspace\psi_{2},\thinspace\psi_{3})$. The fermions get masses mainly via the Yukawa couplings with
a 10-plet scalar fields $H^{(10)}$.
%and at least two 10-plets, $H^{(10)}_u$, $H^{(10)}_d$,
%should be introduced
%to generate different mass hierarchies of the upper and down components of the SM doublets.
Additional non-renormalizable interactions will be added to generate difference
of masses of down quarks and charged leptons.

The hidden sector consists of fermions $S_i$ and bosons $\phi_i$, which are all singlets of $SO(10)$.
In the simplest version three fermionic singlets
%%$(S_{1},\thinspace S_{2},\thinspace S_{3})$
are introduced. Connection between the visible and hidden sectors is established via the portal interaction of $S_i$
and $\psi_{i}$. For this the 16-plet of scalar fields, $H^{(16)}$, should be introduced.

Thus, the Yukawa interactions relevant for generation of fermion masses are
\begin{equation}
{\cal L} \supset  y_{ij}^{v}\psi_{i}\psi_{j}H^{(10)}+y_{ij}^{p}\psi_{i}S_{j}H^{(16)}
+ h_{ij} S_{i} S_{j} \phi_{ij},
\label{eq:gut-15}
\end{equation}
where  $y_{ij}^{v}$,  $y_{ij}^{p}$  and $h_{ij}$ are the Yukawa coupling constants of the
visible, portal and  hidden sectors correspondingly.\\
%[[put equation before description ? ]]\\

{\it 2. Double seesaw.} After the scalar fields develop VEV's,
the visible sector interactions generate the Dirac mass matrices $m_D = y^v_{ij} \langle H^{(10)} \rangle $
at the electroweak (EW) scale.
%[[distinguish for upper and down fermions]]
The portal interactions produce the matrix $M_{RS} = y_{ij}^{p} \langle H^{(16)} \rangle$ which
mixes the RH neutrinos with the singlet fermions.\footnote{A linear seesaw contribution
$m_{LS}$ (see \cite{Ludl:2015tha}) can also be generated,
but in this framework its contribution is negligibly small.}
Flavon VEV's $\langle \phi_{ij} \rangle$ generate the mass matrix of singlets
$M_{S ij} =  h_{ij} \langle \phi_{ij} \rangle$.
Consequently, the total mass matrix of neutral leptons in the basis
$(\nu_{L},\thinspace\nu^c_{L},\thinspace S)$ (here $\nu^c_{L} \equiv (\nu_R)^c $)  becomes
\begin{equation}
{\cal M} = -\frac{1}{2}
\left(\begin{array}{ccc}
0 & m_{D} & 0\\
m_{D}^{T} & 0 & M_{RS}\\
0 & M_{RS}^{T} & M_{S}
\end{array}\right),
\label{eq:gut-7}
\end{equation}
which is  the mass matrix of the double seesaw mechanism \cite{Mohapatra:1986aw}.
It leads to  the Majorana masses of the RH neutrinos
$\nu_{R}$
\begin{equation}
M_{R}=-M_{RS}M_{S}^{-1}M_{RS}^{T},
\label{eq:gut-8}
\end{equation}
and the mass matrix of light neutrinos $\nu_L$
\begin{equation}
m_{\nu} = m_{D}\left(M_{RS}^{-1}\right)^{T}M_{S}M_{RS}^{-1}m_{D}^{T}.
\label{eq:gut-12}
\end{equation}

The Higgs multiplet $H^{(16)}$ breaks the  GUT symmetry,  so that
the natural scale of the largest portal mass term is
$M_{RS} \sim M_{\rm{GUT}} =  {\cal O}(10^{16}{\rm \ GeV})$.
The singlets $S$, which are not protected by $SO(10)$, may have masses of higher scales,
say the string-Planck scale, $M_S \sim M_{\rm{Pl}} \sim (10^{18}- 10^{19})$ GeV.
%%
%%Here $m_{D}$ is the Dirac mass matrix, which in $SO(10)$ models
%%is generally at the same order of the top-type quark mass matrix i.e.
%%${\cal O}(10^{2}{\rm \ GeV})$.
%%The mass matrix $M_{RS}$, which mixes
%%right-handed neutrinos $\nu_{R}$ with the singlets $S$, should be
%%at the GUT scale ${\cal O}(10^{16}{\rm \ GeV})$ because it is generated
%%by $\langle H^{(16)}\rangle$, while the mass matrix of the singlets
%%$M_{S}$ can be even higher, e.g. $10^{18}$ GeV. At the GUT scale,
%%the singlets can be integrated out,
In this case  $M_R =   M_{\rm{GUT}}^2/M_{\rm{Pl}} =  {\cal O}(10^{14}{\rm \ GeV})$, which produces
correct scale of light neutrino masses $m_{\nu}=  {\cal O}(0.1{\rm \ eV})$.
This coincidence can be considered as another support of the framework.\\

{\it 3.  Intrinsic symmetries.} Both the visible and  hidden
sector interactions (\ref{eq:gut-15})
have the built-in $\mathbb{Z}_{2}\times\mathbb{Z}_{2}\times\mathbb{Z}_{2}$ flavor
symmetries
%%, which can be seen as a direct generalization of the similar
%%conclusion in
~\cite{Lam:2007qc,Lam:2008rs,Lam:2008sh,Lam:2014kga}.
The symmetries  are related to the Majorana character of interaction ($\psi^T \cdot \psi$,~ $S^T \cdot S$ )
and obvious in the basis where the mass matrices are diagonalized.
In a general basis, the visible interactions are
%%$y_{ij}^{v}\psi_{i}\psi_{j}$
invariant under the transformation
\begin{equation}
\psi\rightarrow T\psi, \, \, \,
T = U_{\psi}{\rm diag}\left[ (-1)^m, \thinspace (- 1)^n,\thinspace  (- 1)^k \right] U_{\psi}^{\dagger},
\label{eq:gut-24}
\end{equation}
where  $m, n, k = 0, 1$,  and $U_{\psi}$ is a unitary matrix that diagonalizes $m_D$.
Different choices of $m, n, k$  generate different
$T$'s, including trivial cases $T = \pm \mathbb{I}$.
These $T$'s form the $\mathbb{Z}_{2}\times\mathbb{Z}_{2}\times\mathbb{Z}_{2}$
group, which can be reduced to $G_V = \mathbb{Z}_{2}\times\mathbb{Z}_{2}$
if generators with the overall negative sign are removed.
Similarly, in the hidden sector the mass terms
$1/2 \left(M_{S}\right)_{ij}S_{i}S_{j}$
%$M_{S}_{ij}S_{i}S_{j}$
are invariant under
\begin{equation}
S \rightarrow RS,\,  \,
R = U_{S}{\rm diag}\left[(- 1)^m, \thinspace (- 1)^n,\thinspace  (- 1)^k \right] U_{S}^{\dagger},
\label{eq:gut-25}
\end{equation}
where $U_{S}$ is a unitary matrix that diagonalizes $M_{S}$.
So, the hidden sector has another $\mathbb{Z}_{2}\times\mathbb{Z}_{2}$ symmetry, which is denoted as $G_H$ henceforth. \\
%%To distinguish between them, we will henceforth denote the
%%$\mathbb{Z}_{2}\times\mathbb{Z}_{2}$ symmetries in the hidden and

{\it 4. Screening.} The condition
\begin{equation}
m_{D}\propto M_{RS}^{T}
\label{eq:screening}
\end{equation}
leads, as follows from Eq.~(\ref{eq:gut-12}), to
\begin{equation}
m_{\nu}\propto M_{S},
\label{eq:gut-14}
\end{equation}
{\it i.e.} to ``screening'' (cancellation) of the Dirac structures, and consequently,  to
the same structure of the mass matrices of light neutrinos and  the heavy singlets \cite{Lindner:2005pk}.
Consequently the light neutrinos and the heavy singlets have
the same mixing. The RGE effects do not destroy the cancellation \cite{Lindner:2005pk}.
%\footnote{The RGE effect do not destroy the cancellation.%, and can lead to small corrections to our results.
%}

The screening condition (\ref{eq:screening}) can be a consequence of
further unification, e.g.  embedding of $\psi$ and $S$ into
the 27-dimensional representation of $E_{6}$-multiplet.
It can be a remnant of $E_{6}$ symmetry which is not
fully realized. In this case $S$ can not be considered as belonging to
the hidden sector.
Another possibility is a common flavor symmetry acting in the visible and portal sectors.
In fact,  it is easy to get
$$
m_{D} M_{RS}^{-1 T} = d,
$$
where $d$ is a diagonal matrix. The Klein symmetry
$\mathbb{Z}_{2}\times\mathbb{Z}_{2}$ with the same charge assignment
for $\phi_i$ and $S_i$ allow to achieve this. If the charges of three components $\psi_1$, $\psi_2$,
$\psi_3$ are different, e.g. $(-, -)$ $(-, +)$, $(+, -)$ and scalar multiplets
$H^{(10)}$,   $H^{(16)}$ have zero charges,  the matrices $m_{D}$ and $M_{RS}$
are both diagonal \cite{Ludl:2015tha}. However, additional symmetry
should be introduced to make ratios of the element in $m_{D}$
and $M_{RS}$ to be equal, so that $d = \mathbb{I}$.
%In other words,
%$\mathbb{Z}_{2}\times\mathbb{Z}_{2}$ allows to make both matrices diagonal
%but diagonal elements of $m_{D}$ and  $M_{RS}$ are in general different.
A kind of  permutation symmetry
$\psi \leftrightarrow S$  could be used. \\

{\it 5.  Basis fixing symmetry.}
$G_b = \mathbb{Z}_{2}\times\mathbb{Z}_{2}$  was introduced as the
basis fixing symmetry in all the sectors, thus allowing to communicate
information about mixing from the hidden sector to the visible one.
This symmetry
%%The described $\mathbb{Z}_{2}\times\mathbb{Z}_{2}$ symmetry
leads to diagonal structure of all Dirac mass matrices, even if several
Higgs 10-plets (or other representations) with zero charges are introduced.
This means that no CKM-mixing is generated:  $U_{{\rm CKM}}= \mathbb{I}$.
Therefore  generation of CKM mixing would require  breaking
of the $\mathbb{Z}_{2}\times\mathbb{Z}_{2}$ symmetry.
$G_b$ can be identified with $G_V$.
To generate mixing $U_0$,  $G_b$ should be broken in the hidden sector. \\
%%so that intrinsic (unbroken) symmetry  $G_H$ differs from  $G_b = G_V$.
%In other words, basis in which $M_S$ is diagonal differ from the basis where
%$m_D$ is diagonal.
%Bases of $G_b = G_V$ and $G_H$ are different,
%%that the bases of $(\mathbb{Z}_{2}\times\mathbb{Z}_{2})^{H}$ and
%%$(\mathbb{Z}_{2}\times\mathbb{Z}_{2})^{v}$,
%%The bases in which the generators of these groups are diagonal are different,
%%{\it i.e.}, $U_{S}$ and $U_{\psi}$, differ from each other.
%The breaking leads to difference of bases of intrinsic symmetries in
%the visible  and hidden sectors.

{\it 6. Flavons. }
The key element of the framework is that scalars  of the hidden sector $\phi$
do carry non-trivial $G_b$ charges, in contrast to
$H^{(10)}$  and $H^{(16)}$.  When $\phi_i$ get  VEVs,
 $G_b$  is spontaneously broken in
the hidden sector. This leads to non-diagonal matrix $M_S$, and consequently, to
mixing of singlets $S_i$. It is this mixing that generates
the matrix $U_0$.
The Klein  (abelian) symmetry is not enough to obtain special form of $M_S$ and
consequently, $U_0$ like TBM.
For this non-abelian symmetry should be introduced in the hidden sector.

%%(The mass matrices of charged fermions, in
%%the basic framework described by Eq.~(\ref{eq:gut-15}),
%%are all proportional to $y_{ij}^{v}$, leading to $U_{{\rm CKM}}=1$ (generating small
%%mixing in the CKM mixing will be discussed later). )\\

%%visible sector as  $(\mathbb{Z}_{2}\times\mathbb{Z}_{2})^{H}$ and
%%$(\mathbb{Z}_{2}\times\mathbb{Z}_{2})^{v}$ repectively.
%[[one consider the case of non-trivial portal which can also contribute to mixing]]
%
%
%If $m_\nu = M_s$ (i.e. portal doe not contribute to
%mixing) this connecting matrix  gives the lepton mixing matrix $U_0$.

\section{$U_0$ from  the residual symmetries of visible and hidden sectors \label{sec:symmetry}}
%%%%%%%%%%%%%%%%%%%%%%%%%%%%%%%%%%%%%%%%%%%%%%%%%%%%%%%%%%%%%%%%%%%%%%%%%%%%%%%%%%

In the case of complete screening,  the portal interactions
do not influence the mixing and one can immediately discuss the bases of the visible and
hidden sectors. Then the mixing can be understood as a relation
between the bases in which  generators of $G_V$ and $G_H$ have diagonal forms.
%%[[$G_V$ is realized in the basis of $\psi$
%%whereas $G_H$ in the basis  of $S$ connection via portal]].
According to (\ref{eq:gut-24}) and (\ref{eq:gut-25}), these two basis should be connected by
$
U_{0} = U_{\psi}^{\dagger} U_{S}.
$
Thus, we start with common basis fixing symmetry  $G_b$ in all the sectors
and then to promote $G_b$ to a larger non-abelian group $G_f$
in the hidden sector.
Then breaking of $G_f$ should be arranged in such a way
that $G_b \in G_f$ is broken and another intrinsic
unbroken $G_H =  (\mathbb{Z}_{2}\times\mathbb{Z}_{2})^{H}$ symmetry is realized.

%(or between the corresponding two bases
%where the corresponding matrices are diagonal respectively).
%%Embedding of $G_V$ and $G_H$
%%into a unique flavor group  $G_f$ can  produce such a connection:
%In other words,  $G_f$ can be introduced in all the sectors.
%Then $G_f$ is broken down to residual symmetry group $G_V$ in the visible sector,
%and to intrinsic group $G_H$ in the hidden sector:

%%Another point of view is that
%%to transmit information from hidden sector to the visible one
%%we should establish connections between  $G_V$ and $G_H$.
%%Embedding of
%%$G_V$ and $G_H$ are embedded into a unique flavor group  $G_f$:

One can view this procedure as the residual symmetry approach:
$G_H$ and $G_V$ are embedded into a unified flavor group
\begin{equation}
G_{f}\supset G_H, \,  G_V.
%%(\mathbb{Z}_{2}\times\mathbb{Z}_{2})^{v},\ (\mathbb{Z}_{2}\times\mathbb{Z}_{2})^{H}.
\label{eq:gut-26}
\end{equation}
This embedding ensures that information about mixing
from the hidden sector is transmitted to the visible sector.
Then $G_f$ is broken explicitly down to $G_V$ in the
visible (low mass scale) sector and it is broken down to $G_H$ spontaneously
in the hidden sector.\footnote{One can consider also spontaneous symmetry breaking in the
visible and portal sectors, but this would introduce
further complication of the model}

%Then break $G_f$ down to $G_H$  (which differs from $G_b$) in the hidden sector.
%So $G_V$  and $G_H$ appear as residual symmetries.
This is similar to the usual residual symmetry approach
\cite{Lam:2007qc,Lam:2008rs,Lam:2008sh,Lam:2014kga} when instead of mass matrices of the charged
leptons and neutrinos we use the mass matrices of $\psi$ and $S$.
Here the residual symmetries operate at different
energy scales: the GUT-scale and the Planck scale.

An important feature  is that  in both sectors the residual symmetries are
given by the Klein groups:
\begin{equation}
G_V = (\mathbb{Z}_{2}\times\mathbb{Z}_{2})^{V}, ~~~~
G_H =  (\mathbb{Z}_{2}\times\mathbb{Z}_{2})^{H}.
\label{eq:residgr}
\end{equation}
Embedding of two Klein groups into a finite group
%%$\mathbb{Z}_{2}\times\mathbb{Z}_{2}$
and its consequences for mixing have been explored in
\cite{Bajc:2016eiw}.
The only difference is that in \cite{Bajc:2016eiw} the results of embedding were  applied  to
the relative matrix between interactions with
Higgs 10-plet and Higgs 126-plet, while here we deal with the relative rotation
between the mass basis  generated by Higgs 10-plet
(visible sector which coincides with matrix of portal)  and the mass basis
generated by Higgs singlets in the hidden sector.
In what follows, we briefly remind the  important points and present the  main results.

In a 3-dimensional irreducible representation  of $G_{f}$, the elements
%%in $(\mathbb{Z}_{2}\times\mathbb{Z}_{2})^{v}$
%%and $(\mathbb{Z}_{2}\times\mathbb{Z}_{2})^{H}$ are given
$T \in G_V $ and $R \in G_H$ in Eq.~(\ref{eq:gut-24}) and Eq.~(\ref{eq:gut-25})
 with positive determinants  can be written as
%. In the basis where $T$ are diagonal the elements
%%Since one can always choose a proper basis for this
%%3-dimensional irrep, we will work in the $T$-diagonal basis.
%%For
%%the convenience of later use, we explicitly write down all the nontrivial
%%(i.e. not the identity matrix) elements
%%in $(\mathbb{Z}_{2}\times\mathbb{Z}_{2})^{v}$ and $(\mathbb{Z}_{2}\times\mathbb{Z}_{2})^{H}$:
%%\begin{equation}
%%T_{1,\thinspace2,\thinspace3}\in(\mathbb{Z}_{2}\times\mathbb{Z}_{2})^{v},\
%%R_{1,\thinspace2,\thinspace3}\in(\mathbb{Z}_{2}
%%\times\mathbb{Z}_{2})^{H},\label{eq:gut-79}
%%\end{equation}
\begin{equation}
T_{1}=\left(\begin{array}{ccc}
1\\
 & -1\\
 &  & -1
\end{array}\right),\ ~~ T_{2}=\left(\begin{array}{ccc}
-1\\
 & 1\\
 &  & -1
\end{array}\right),\ ~~ T_{3}=\left(\begin{array}{ccc}
-1\\
 & -1\\
 &  & 1
\end{array}\right),\label{eq:gut-80}
\end{equation}
\begin{equation}
R_{1}=U_{0}\left(\begin{array}{ccc}
1\\
 & -1\\
 &  & -1
\end{array}\right)U_{0}^{\dagger},\ ~~ R_{2}=U_{0}\left(\begin{array}{ccc}
-1\\
 & 1\\
 &  & -1
\end{array}\right)U_{0}^{\dagger},\ ~~ R_{3}=U_{0}\left(\begin{array}{ccc}
-1\\
 & -1\\
 &  & 1
\end{array}\right)U_{0}^{\dagger},
\label{eq:gut-81-1}
\end{equation}
where we use the basis in which $T$ are diagonal.
%%where $U_{0}=U_{\psi}^{\dagger}U_{S}$.
By definition the  group $G_{f}$
should contain all these elements as well as their products.
%%such
%%as $T_{1}R_{2}$, $T_{2}R_{3}$, $R_{1}T_{1}R_{2}$, etc.
Furthermore, since  $G_{f}$ is a finite group, any product of $T$ and $R$ should have a finite
order:

\begin{equation}
W_{ij}^p \equiv   \left(T_{i}R_{j}\right)^{p} =
\left(T_{i} U_0 R_{j}^d U^{\dagger}_0 \right)^{p}=\mathbb{I}, \ \ (i,\thinspace j=1,\thinspace 2,\thinspace3),
\label{eq:gut-83}
\end{equation}
where $p$ is a positive integer. This is the symmetry group condition \cite{Bajc:2016eiw}
which determines the $i$-$j$ element of $U_0$:
%%%%%%%%%%%%%%%%%%%%%%%%%%%%%%%%%%%%%%%%%%%%%%%%%%%%%%%%%%%%%%%%%%
\begin{equation}
|(U_0)_{ij}|^2 = \cos^2 \frac{\pi n}{p},
\label{eq:gut-27}
\end{equation}
where $n$ and $p$ are integers. The detailed derivation of Eq.~(\ref{eq:gut-27})
is presented in the Appendix \ref{sec:cos}.
%%In other words, $\alpha$ is a rational
%%multiple of $\pi$. For our convenience, we will henceforth refer
%%to such angles as \emph{rational angles}.
Using one generator $T_i \in G_V$  and another one
$R_{j} \in  G_H$ fixes the $i$-$j$ element of the matrix $U_0$.
The indices $i, j = 1, 2, 3$ are identified by the positive diagonal elements in Eq.~(\ref{eq:gut-80}) and Eq.~(\ref{eq:gut-81-1}). For instance,  $i = 1$ corresponds  to generators with +1 in the 1-1 position,
{\it etc.}

For each pair of $(i, j)$, it can be its own power $p_{ij}$, and the corresponding
$n_{ij} < p_{ij}$. Furthermore, for fixed  $p_{ij}$, several $n_{ij}$ can exist.
Taking three different symmetry group relations given by one $T$ and three $R$
(or vice versa)
we can fix three elements of row (column) of the mixing matrix and they should
satisfy the unitarity condition.
Using (\ref{eq:gut-27}) we can write the unitarity condition, in general,   as
%[[should it be the same p for all ?]]
\begin{equation}
\cos^{2}\alpha+\cos^{2}\beta+\cos^{2}\gamma=1,
\label{eq:gut-90}
\end{equation}
where $\alpha$, $\beta$ and $\gamma$ are rational numbers of $\pi$.

As we will see,  the unitarity condition in the form (\ref{eq:gut-90}) already strongly
restricts the number of possibilities even before further
applications of the group theory constraints.
Without loss of generality,
we assume that $\cos\alpha\leq\cos\beta\leq\cos\gamma$ and
$0\leq\alpha,\thinspace\beta,\thinspace\gamma\leq90^{\circ}$.
%%Besides, because
%%Eq.~(\ref{eq:gut-90}) is invariant when the three angles change their
%%signs or increase by $180^{\circ}$, we are only concerned with the
%%solutions in the first quadrant:
%\begin{equation}
%0\leq\alpha,\thinspace\beta,\thinspace\gamma\leq90^{\circ}.
%\label{eq:gut-93}
%\end{equation}
Under these assumptions, we only need to consider two cases:   $\cos\alpha=0$
and   $\cos\alpha\neq0$.

%%\vspace{2mm}
%%{\noindent (i) $\cos\alpha=0$, $\alpha = \pi/2$}
%%\vspace{2mm}

If $\cos\alpha$ is zero ($\alpha = \pi/2$),
Eq.~(\ref{eq:gut-90}) reduces to $\cos^{2}\beta+\cos^{2}\gamma=1$,
from which one immediately obtains $\gamma=\pi/2-\beta$. Since  $\beta = \pi q/p$,
we get infinite number of solutions for the angles:
\begin{equation}
(\alpha,\thinspace\beta,\thinspace\gamma) =
\pi \left(\frac{1}{2},~
\frac{q}{p},~ \frac{1}{2}-\frac{q}{p} \right).
%(\cos \alpha,~ \cos\beta, \cos \gamma) =
%\left(0, ~  \cos \pi \frac{q}{p}, ~ \cos \pi \left(\frac{1}{2} -  \frac{q}{p} \right) \right)
\label{eq:gut-94}
\end{equation}

For $\cos\alpha\neq0$ (all cosines are non-zero)
a numerical search %for the solutions of Eq. (\ref{eq:gut-90})
for all rationals $q/p$  with $p \leq 100$
gives only two solutions:
\begin{equation}
(\alpha,\thinspace\beta,\thinspace\gamma)= \pi
\left(\frac{1}{3},~ \frac{1}{3},~\frac{1}{4}\right) ,
\label{eq:gut-95}
\end{equation}
and
\begin{equation}
(\alpha,\thinspace\beta,\thinspace\gamma) =  \pi \left(\frac{2}{5},~
\frac{1}{3},~ \frac{1}{5}\right).
\label{eq:gut-96}
\end{equation}

The elements  of $|U_0|$ which correspond to   (\ref{eq:gut-94}), (\ref{eq:gut-95})
(\ref{eq:gut-96}) are
%
%%In summary, the rational angle solutions of Eq.~(\ref{eq:gut-90})
%%are Eq.~(\ref{eq:gut-94}), Eq.~(\ref{eq:gut-95}) and Eq.~(\ref{eq:gut-96}).
%%Therefore the columns of $\left|U_{0}\right|$ can only be the following
%%vectors:
\begin{equation}
v_{1} \equiv \left( 0,~ \cos\frac{q}{p}\pi, ~ \sin\frac{q}{p}\pi \right),  ~~~~
v_{2} \equiv \left(\frac{ 1}{\sqrt{2}}, ~ \frac{1}{2},~ \frac{1}{2} \right),  ~~~~
v_{3} \equiv \left(\frac{\sqrt{5} +1}{4}, ~\frac{1}{2}, ~ \frac{\sqrt{5} - 1}{4} \right).
\label{eq:gut-98}
\end{equation}
%%
%%or other vectors that can be converted to the above vectors by permutationsof their entries.
%%\begin{equation}
%%v_{1}\equiv\left(\begin{array}{c} 0\\
%%\cos\frac{q}{p}\pi\\
%%\sin\frac{q}{p}\pi
%%\end{array}\right),\ v_{2}\equiv\left(\begin{array}{c}
%%1/\sqrt{2}\\
%%1/2\\
%%1/2
%%\end{array}\right),\ v_{3}\equiv\left(\begin{array}{c}
%%(1+\sqrt{5})/4\\
%%1/2\\
%%(\sqrt{5}-1)/4
%%\end{array}\right),
%%\label{eq:gut-98}
%%\end{equation}
%%%%%%%%%%%%%%%%%%%%%%%%%%%%%%%%%%%%%%%%%%%%%%%%%%%%%%%%
%%
The last solution in Eq.~(\ref{eq:gut-98}) can be expressed in terms of the golden ratio,
\begin{equation}
\varphi\equiv\frac{1}{2}\left(1+\sqrt{5}\right)\approx1.618,
\label{eq:gut-97}
\end{equation}
$v_{3} =
\frac{1}{2}(\varphi,\thinspace1,\thinspace\varphi^{-1})^{T}$.
%It has been applied for  neutrino flavor symmetries before \cite{}.
The first solution in (\ref{eq:gut-98})
has two interesting possibilities:
\begin{equation}
v_{1a} = (1,~ 0,~ 0), ~~~~v_{1b} =
\left(\frac{1}{\sqrt{2}},~ \frac{1}{\sqrt{2}}, ~ 0 \right).
\label{eq:special}
\end{equation}

%%\vspace{2mm}

%%{\noindent \bf \large Rational mixing matrices and finite groups}

%%\vspace{2mm}

%%Now we come back to Eq.~(\ref{eq:gut-81}) where all the $\alpha_{ij}$'s
%%should be rational angles. We call such a unitary matrix as a rational
%%mixing matrix. With the previous conclusion, it is quite straightforward

Now using the vectors in Eq.~(\ref{eq:gut-98}) as building blocks,  we can
construct  complete mixing matrices. There is a freedom to take
$v_i$ as rows or columns of the matrix  and also to permute elements
within $v_i$. Not all combinations are allowed by unitarity.
If the unitarity is satisfied for column, one should arrange the elements in each column so that it is
satisfied for rows as well.
Clearly, using three times the same column with permuted elements
will automatically satisfy the unitarity condition for whole the matrix.

Let us consider first that at least one of the columns of $|U_{0}|$
is in the form of $v_{1}$ in Eq.~(\ref{eq:gut-98}), which means
that $U_{0}$ has at least one zero entry.
In general, one can prove\footnote{
The proof is straightforward enumeration. The number of zeros
can not be larger than 6 because it implies that at most
two elements of $U_0$ can be non-zero, which is impossible for a unitary matrix.
If there are five zeros, i.e., four elements are non-zero,
then the orthogonality of rows/columns requires one of the four elements to be zero.
Likewise, one can check that $U_0$ with two or three zeros have the same problem.
}
that the number of zero entries in a $3\times3$ unitary matrix can only be 1 or 4 or 6.
The matrix with  4 zeros constructed of $v_1$ is
\begin{equation}
|U_{0}| = \left(\begin{array}{ccc}
1 & 0 & 0\\
0 & \cos\frac{q}{p}\pi & \sin\frac{q}{p}\pi\\
0 & \sin\frac{q}{p}\pi & \cos\frac{q}{p}\pi
\end{array}\right).
\label{eq:gut-99}
\end{equation}
The case of 6 zeros  corresponds
to  $|U_{0}|=I$.

For the case of single zero,  we take $v_1$ while the other columns can not contain
zero entries, and therefore they have to be of the form  $v_{2}$ or $v_{3}$.
The conclusion about the columns of $\left|U_{0}\right|$
also hold for the rows, which implies that the rows consist of one
$v_{1}^{T}$, and two $v_{2}^{T}$ or $v_{3}^{T}$. As a result, the
two non-zero elements in $v_{1}$ have to be $1/2$, $1/\sqrt{2}$, $\varphi/2$,
or $\varphi^{-1}/2$. The squared sum of the two non-zero elements should
be 1, leaving only one option: $(0, 1/\sqrt{2},\ 1/\sqrt{2})$. Therefore,
in the case of 1 zero, we have:
\begin{equation}
|U_{0}|=\left(\begin{array}{ccc}
1/\sqrt{2} & 1/\sqrt{2} & 0\\
1/2 & 1/2 & 1/\sqrt{2}\\
1/2 & 1/2 & 1/\sqrt{2}
\end{array}\right),
\label{eq:gut-100}
\end{equation}
which coincides with the BM mixing matrix %\cite{Vissani:1997,Barger:1998ta,He:2012yt}
\cite{Barger:1998ta,Mohapatra:1998ka,Nomura:1998gm,
Frampton:2004ud,Altarelli:2009gn,He:2012yt}.
There is no other possibilities with $v_1$.

Next, let us consider  $|U_{0}|$ constructed from $v_{2}$ , or $v_{3}$, or $v_{2}$
and $v_{3}$ together. There is only one solution in each case:
\begin{equation}
|U_{0}|=  \frac{1}{2} \left(\begin{array}{ccc}
\sqrt{2} & 1 & 1\\
1 & \sqrt{2} & 1\\
1 & 1 & \sqrt{2}
\end{array}\right),\ ~~
\frac{1}{2} \left(\begin{array}{ccc}
\varphi  & 1 &  \varphi^{-1}\\
1 & \varphi^{-1} & \varphi\\
\varphi^{-1} & \varphi  & 1
\end{array}\right),\ ~~
\frac{1}{2}\left(\begin{array}{ccc}
\varphi^{-1}   & 1 & \varphi \\
1 & \sqrt{2} & 1\\
\varphi & 1 & \varphi^{-1}
\end{array}\right).
\label{eq:gut-101}
\end{equation}

%The symmetry group conditions fix the moduli of the matrix elements, Eq.~(\ref{eq:gut-86}).
The symmetry group condition gives the moduli of elements,   $|U_0|$.
%%while the phases in $U_0$ is still unknown.
To reconstruct $U_0$ completely one needs to find the phases of elements
which ensure orthogonality of the rows and columns
in $U_{0}$. Without loss of generality, we assume  that $u_{11}$,
$u_{12}$, $u_{13}$, $u_{23}$, and $u_{33}$ are real. Then  the orthogonality
of the columns gives
\begin{equation}
u_{11}u_{13}+|u_{21}|e^{i\phi_{21}}u_{23}+|u_{31}|e^{i\phi_{31}}u_{33}=0,
\label{eq:gut-102}
\end{equation}
\begin{equation}
u_{12}u_{13}+|u_{22}|e^{i\phi_{22}}u_{23}+|u_{32}|e^{i\phi_{32}}u_{33}=0,
\label{eq:gut-103}
\end{equation}
where $\phi_{ij}\equiv\arg(u_{ij})$.
Using the graphic representation of the  equalities (\ref{eq:gut-102}, \ref{eq:gut-103})
(i.e., the unitarity triangles) we obtain:
\begin{equation}
\cos\phi_{2j}=\frac{|u_{3j}|^{2}u_{33}^{2}-u_{1j}^{2}u_{13}^{2}-
|u_{2j}|^{2}u_{23}^{2}}{2u_{1j}u_{13}|u_{2j}|u_{23}},~~\
\cos\phi_{3j}=\frac{|u_{2j}|^{2}u_{23}^{2}-u_{1j}^{2}u_{13}^{2}-
|u_{3j}|^{2}u_{33}^{2}}{2u_{1j}u_{13}|u_{3j}|u_{33}},\ (j=1,2).
\label{eq:gut-104}
\end{equation}

%Using Eqs.~(\ref{eq:gut-104})
Consequently, the phases for the three matrices in Eq.~(\ref{eq:gut-101}) equal
\[
{\rm arg}(U_{0})=\left(\begin{array}{ccc}
0 & 0 & 0\\
\arccos\frac{-1}{2\sqrt{2}} & -\arccos\frac{-3}{4} & 0\\
-\arccos\frac{-3}{4} & \arccos\frac{-1}{2\sqrt{2}} & 0
\end{array}\right),\ \left(\begin{array}{ccc}
0 & 0 & 0\\
\pi & -\pi & 0\\
0 & \pi & 0
\end{array}\right),\ \left(\begin{array}{ccc}
0 & 0 & 0\\
2\pi/3 & -\arccos\left(\sqrt{2}\left(\sqrt{5}-3\right)\right)^{-1} & 0\\
-2\pi/3 & 2\pi/3 & 0
\end{array}\right).
\]
%(The CP  phase in the standard parametrization or the phase matrix of  $U_0$
%after $V_{\rm{CKM}}$  in Eq. (\ref{eq:gut-45}) is model-dependent.)
The matrices in Eqs.~(\ref{eq:gut-99}) and (\ref{eq:gut-100}),
contain zero mixing angles and therefore
the phases can be removed by rephasing.
Orthogonality in Eq.~(\ref{eq:gut-99}) and Eq.~(\ref{eq:gut-100}) can be achieved
by adding minus signs,
e.g.  to 2-1, 3-1, 2-3 elements in Eq.~(\ref{eq:gut-100})
and to 3-2 element in Eq.~(\ref{eq:gut-99}).
%An interesting special case of the matrix (\ref{eq:gut-99}) is
%\begin{equation}
%\left(\begin{array}{ccc}
%1 & 0 & 0\\
%0 & 1/\sqrt{2} & 1/\sqrt{2}\\
%0 & -1/\sqrt{2} & 1/\sqrt{2}
%\end{array}\right).
%\label{eq:gut-105}
%\end{equation}
%[[Comparison with Fonseca. ]]\xj{to be added later}
%\xxj the text below should be refined.
The %complete  forms of the
mixing matrices including the phases are
summarized in the Table ~\ref{tab:rational}.
%%We denoted  these mixing matrices by the rational angles
%%(in degrees) that appear in the matrices for non-block diagonal matrices
%%or the rational number $q/p$ for block diagonal matrices. For example,
%%the mixing 90-60-45 referes to the BM mixing and the 1/2 mixing is
%However, a significant difference should be noticed.
%in a similar way to our work.

%%%%%%%%%%%%%%%%%%%%%%%%%%%%%%%%%%%%%%%%%%%%%%%%%%%%%%%%%%%%%%%%%%%%%%%%%%%%%%%%%%%%
\begin{table*}
\caption{\label{tab:rational} All rational mixing matrices with $p\leq100$
and the corresponding finite groups.}

\begin{ruledtabular}
\begin{tabular}{ccccc}
Matrix &  $|U_{0}|$ & $U_{0}$ & $(T_{i}R_{j})^{p}=1$ & Group\tabularnewline
\hline
$U_{q/p}$ &  $\cos\left(\begin{array}{ccc}
0 & \frac{\pi}{2}  &  \frac{\pi}{2}\\
\frac{\pi}{2} & \frac{q}{p}\pi & \frac{\pi}{2}-\frac{q}{p}\pi\\
\frac{\pi}{2} & \frac{\pi}{2}-\frac{q}{p}\pi & \frac{q}{p}\pi
\end{array}\right)$ & $\left(\begin{array}{ccc}
1 & 0 & 0\\
0 & \cos\frac{q}{p}\pi & \sin\frac{q}{p}\pi\\
0 & -\sin\frac{q}{p}\pi & \cos\frac{q}{p}\pi
\end{array}\right)$ &$\left[\begin{array}{c}
(T_{1}R_{1})^{1}\\
(T_{1}R_{2})^{2}\\
(T_{2}R_{1})^{2}\\
(T_{2}R_{2})^{p}
\end{array}\right]=1$ & $D_{p}$\tabularnewline
$U_{\rm{BM}}$ &  $\cos\left(\begin{array}{ccc}
45^{\circ} & 45^{\circ} & 90^{\circ}\\
60^{\circ} & 60^{\circ} & 45^{\circ}\\
60^{\circ} & 60^{\circ} & 45^{\circ}
\end{array}\right)$ & $\left(\begin{array}{ccc}
1\left/\sqrt{2}\right. & 1\left/\sqrt{2}\right. & 0\\
-1/2 & 1/2 & -1\left/\sqrt{2}\right.\\
-1/2 & 1/2 & 1\left/\sqrt{2}\right.
\end{array}\right)$ & $\left[\begin{array}{c}
(T_{1}R_{1})^{4}\\
(T_{1}R_{2})^{4}\\
(T_{2}R_{1})^{3}\\
(T_{2}R_{2})^{3}
\end{array}\right]=1$ & $S_{4}$\tabularnewline
$U_{v2}$ &  $\cos\left(\begin{array}{ccc}
45{}^{\circ} & 60{}^{\circ} & 60{}^{\circ}\\
60{}^{\circ} & 45{}^{\circ} & 60{}^{\circ}\\
60{}^{\circ} & 60{}^{\circ} & 45{}^{\circ}
\end{array}\right)$ & $\left(\begin{array}{ccc}
\frac{1}{\sqrt{2}}    & \frac{1}{2} & \frac{1}{2} \\
\frac{1}{2}  & -\frac{\sqrt{7}i+3}{4\sqrt{2}}   &  \frac{\sqrt{7}i-1}{4\sqrt{2}} \\
\frac{1}{2}  & \frac{\sqrt{7}i-1}{4\sqrt{2}} &    -\frac{\sqrt{7}i+3}{4\sqrt{2}}
\end{array}\right)$ & $\left[\begin{array}{c}
(T_{1}R_{1})^{3}\\
(T_{1}R_{2})^{3}\\
(T_{2}R_{1})^{3}\\
(T_{2}R_{2})^{4}
\end{array}\right]=1$ & $PSL(3,2)$\tabularnewline
$U_{GR}$ &  $\cos\left(\begin{array}{ccc}
36{}^{\circ} & 60{}^{\circ} & 72{}^{\circ}\\
60{}^{\circ} & 72{}^{\circ} & 36{}^{\circ}\\
72{}^{\circ} & 36{}^{\circ} & 60{}^{\circ}
\end{array}\right)$ & $\left(\begin{array}{ccc}
\varphi/2  & 1/2    & \left.\varphi^{-1}\right/2 \\
1/2 & \left.-\varphi^{-1}\right/2 & -\varphi/2\\
\left.\varphi^{-1}\right/2 & -\varphi/2 & 1/2
\end{array}\right)$ & $\left[\begin{array}{c}
(T_{1}R_{1})^{5}\\
(T_{1}R_{2})^{3}\\
(T_{2}R_{1})^{5}\\
(T_{2}R_{2})^{5}
\end{array}\right]=1$ & $A_{5}$\tabularnewline
$U_{GR-v2}$ &  $\cos\left(\begin{array}{ccc}
36{}^{\circ} & 60{}^{\circ} & 72{}^{\circ}\\
60{}^{\circ} & 45{}^{\circ} & 60{}^{\circ}\\
72{}^{\circ} & 60{}^{\circ} & 36{}^{\circ}
\end{array}\right)$ & $\left(\begin{array}{ccc}
\varphi/2    &  1/2    & \left.\varphi^{-1}\right/2   \\
1/2  & -\frac{2+i\sqrt{3}+\sqrt{5}}{4\varphi} & \frac{1}{4}\left(\sqrt{3}i-1\right) \\
\left.\varphi^{-1}\right/2  & \frac{1}{4}\left(\sqrt{3}i-1\right) &  -\frac{\varphi}{4}\left(\sqrt{3}i+1\right)
\end{array}\right)$ & $\left[\begin{array}{c}
(T_{1}R_{1})^{5}\\
(T_{1}R_{2})^{3}\\
(T_{2}R_{1})^{3}\\
(T_{2}R_{2})^{4}
\end{array}\right]=1$ & $C_{3}\cdot A_{6}$\tabularnewline
\end{tabular}\end{ruledtabular}

\end{table*}
%%%%%%%%%%%%%%%%%%%%%%%%%%%%%%%%%%%%%%%%%%%%%%%%%%%%%%%%%%%%%%%%%%%%

Notice that till now we  used only a general form of matrix elements
as cosines of rational numbers of $\pi$ (\ref{eq:gut-27})  and
the unitarity.  The symmetry group condition is necessary but not sufficient one for
embedding into the finite group.
Still one should check that embedding is possible from the group theory point of view.
 Using $U_{0}$ given in the Table ~\ref{tab:rational},
we can find the corresponding generators $R_i$  according
to  Eq.~(\ref{eq:gut-81-1}), and check the
group presentations.
%Together with  $T_{i}$ we can construct group by multiplying them with each other (also with the new generated elements from the multiplication).
% .....\xj{to be added later}
%(5th column of the Table)
%%and identify the corresponding minimal finite group (6th column).
%%Combined with $T_{1}$, $T_{2}$, $T_{3}$, they can generate the
%%minimal finite group that contains them.
Then using the GAP program \cite{GAP4},
we identify  all the corresponding finite groups (see the last column of
Table~\ref{tab:rational}). Notice that the block diagonal matrices $U_{q/p}$  are  generated
by a dihedral group $D_{p}$, of with the order $p$ determined
by the denominator of the rational angle.
The matrices $U_{\rm{BM}}$, $U_{v2}$, and $U_{GR}$
are generated by the groups $S_{4}$, $PSL(3,2)$ and $A_{5}$, of orders
24, 168 and 60 correspondingly.
It has been well known (see, e.g., \cite{Altarelli:2009gn,He:2012yt}) that the $S_4$ symmetry can be used to obtain $U_{\rm{BM}}$.
The last and most complicated matrix
$U_{GR-v2}$ can be obtained
in a 1080-order group, which is a non-split extension of $A_{6}$
by $C_{3}$, denoted as $C_{3}\cdot A_{6}$ in the GAP classification system.

Reconstructing finite groups from the residual symmetries have been studied
Ref.~\cite{Fonseca:2014koa} using theorems on sums of roots of unity,
which is technically similar to the trace approach formulated in
Eq.~(\ref{eq:gut-85}). Some of the finite groups presented in the Table ~\ref{tab:rational}
(e.g. $S_4$, $A_5$) are the same as those found in \cite{Fonseca:2014koa}.
However, one should note that Ref.~\cite{Fonseca:2014koa} sets
a finite order of $T^{\dagger}RTR$, while in  Eq.~(\ref{eq:gut-83})
we use order of $TR$. Consequently,  we obtain some additional  groups such as $PSL(3,2)$ and $C_{3}\cdot A_{6}$.

In summary,  the BM mixing matrix  can be obtained for $U_0$ in our approach.
Also the matrices $U_{v2}$ (constructed with columns $v_2, v_2, v_2$)
and $U_{GR} (v_3, v_3, v_3)$
can be of the phenomenological interest once certain corrections are taken into account.
The matrix $U_{q/p}$ %in the form (\ref{eq:gut-105})
can be considered for the 2-3  mixing
if the 1-2  mixing is generated,  e.g. from the portal interactions.
Notice that the TBM mixing can not be obtained in this framework.
This is because we require that the residual symmetries should be $\mathbb{Z}_{2}\times\mathbb{Z}_{2}$ while TBM actually needs a $\mathbb{Z}_{3}$ residual symmetry. Without the requirement of $\mathbb{Z}_{2}\times\mathbb{Z}_{2}$, TBM may be obtained in $SO(10)$ frameworks---see e.g. \cite{BhupalDev:2012nm}.

%%%%%%%%%%%%%%%%%%%%%%%%%%%%%%%%%%%%%%%%%%%%%%%%%%%%%%%%%%%%%%%%%%%%%
\section{Bimaximal Mixing from the hidden sector \label{sec:BM}}
%%%%%%%%%%%%%%%%%%%%%%%%%%%%%%%%%%%%%%%%%%%%%%%%%%%%%%%%%%%%%%%%%%%%%%%

 %symmetry group condition considered in the previous section.
%%We use  $S_{4}$  flavor group  as the covering symmetry
%%(which it seems the unique possibility?)

%%The visible and portal sectors are flavor diagonal and respect the same residual symmetry

%%%%%tab2%%%%%%%%%%%%%%%%%%%%%%%%%%%%%%%%%%%%%%%%%%%%%%%%%%%%%%%%%%%%%%%%%%%%
\begin{table}[h]
\centering
\caption{\label{tab:Particles}Field content of the model and symmetry assignments.}
\begin{tabular*}{16cm}{@{\extracolsep{\fill}}cccccccc}
\hline
\hline
 & $\psi$ & $S$ & $H^{(10)}$ & $H^{(16)}$ & $\eta$ & $\xi$ & $\phi$\tabularnewline
\hline
type & fermion & fermion & scalar & scalar & scalar & scalar & scalar\tabularnewline
$SO(10)$ & $\mathbf{16}$ & $\mathbf{1}$ & $\mathbf{10}$ & $\mathbf{16}$ &
$\mathbf{1}$ & $\mathbf{1}$ & $\mathbf{1}$\tabularnewline
$S_{4}$ & $\mathbf{3}$ & $\mathbf{3}$ & $\mathbf{1}$ & $\mathbf{1}$ &
$\mathbf{1}$ & $\mathbf{2}$ & $\mathbf{3}'$\tabularnewline
\hline
\hline
\end{tabular*}

\end{table}
%%%%%%%%%%%%%%%%%%%%%%%%%%%%%%%%%%%%%%%%%%%%%%%%%%%%%%%%%%%%%%%%%%%%%%%%%%

For definiteness we consider generating the BM mixing from the $S_{4}$ embedding of the residual symmetries.
Details of the group $S_{4}$, which has five irreducible representations $\mathbf{1}$,
$\mathbf{1}'$,$\mathbf{2}$, $\mathbf{3}$ and $\mathbf{3}'$,
are given in Appendix \ref{sec:S4}. All the fermions are assigned to the
3-dimensional representation $\mathbf{3}$. The $SO(10)$ Higgs multiplets
$H^{(10)}$, $H^{(16)}$ are flavor singlets. In
contrast,  the Higgs fields in the hidden sector have non-trivial $S_{4}$
assignments, and so the flavor symmetry is broken in this sector spontaneously.
%%addition, we intoduce
%%some flavons (scalar fields with flavor structures) for the hidden
%%sector, which are $SO(10)$ singlets.
The symmetry assignments for the fields are given in the Table~\ref{tab:Particles}.

We assume that in the visible and portal sectors, the $S_{4}$ symmetry
is broken explicitly down to the residual symmetry $(\mathbb{Z}_{2}\times\mathbb{Z}_{2})^{V}$.
%which keeps them diagonal.
The $(\mathbb{Z}_{2}\times\mathbb{Z}_{2})^{V}$
charges of $\psi$ and $S$ are
%%%%%%%%%%%%%%%%%%%%%%%%%%%%%%%%%%%%%%%%%%%%%%%%%%%%%%%%%%%%%%%%%%%%%%%%%%
\[
\begin{array}{ccccccc}
{\rm Fields:}  & \psi_{1},~ S_1 ~~& \psi_{2}, ~S_2 ~~& \psi_{3}, ~S_3 \\
(\mathbb{Z}_{2}\times\mathbb{Z}_{2})^{V}: ~~& (+,-) ~~& (-,+) & (-,-)
\end{array}.
\]
%%%%%%%%%%%%%%%%%%%%%%%%%%%%%%%%%%%%%%%%%%%%%%%%%%%%%%%%%%%%%%%%%%%%%%%%%%%%%%%%%%%%%%%%
%%The scalar fields $H^{(10)}$ and $H^{(16)}$ have zero $\mathbb{Z}_{2}\times\mathbb{Z}_{2}$
%%charges. In the matrix form,
The visible and portal sectors are invariant
under the transformation
\begin{equation}
\psi \rightarrow T_{i} \psi, ~~~~  S \rightarrow T_{i} S, ~~ (i=1,\thinspace 2),
\label{eq:gut-3}
\end{equation}
where  $T_{i}$ are defined in Eq.~(\ref{eq:gut-80}), and
the transformations in Eq.~(\ref{eq:gut-3}) belong to a sub-group of $S_{4}$.

Due to the $(\mathbb{Z}_{2}\times\mathbb{Z}_{2})^{V}$ symmetry
(\ref{eq:gut-3}),
the Yukawa interactions in the visible and portal sectors are flavor-diagonal,
so that the Lagrangian (\ref{eq:gut-15}) reduces to
\begin{equation}
{\cal L}_{\psi}=\sum_{i=1}^{3}\left[y_{i}^{v}\psi_{i}\psi_{i}H^{(10)}+
y_{i}^{p}\psi_{i}S_{i}H^{(16)}\right].
\label{eq:gut}
\end{equation}

In the hidden sector, the Yukawa interactions are
\begin{equation}
{\cal L}_{{\rm hidden}}=y_{ijk}^{\phi}S_{i}S_{j}\phi_{k}+
y_{ijk}^{\xi}S_{i}S_{j}\xi_{k}+y^{\eta}S_{i}S_{i}\eta,\label{eq:gut-5}
\end{equation}
where the Yukawa couplings $y_{ijk}^{\phi}$ and $y_{ijk}^{\xi}$
are  determined by the
$S_{4}$ symmetry.
%%The $S_{4}$ symmetry is broken spontaneously
%%by the flavons $\phi$, $\xi$, and $\eta$.
According to the CG coefficients of $S_{4}$
(see Appendix B), the products of these Yukawa couplings with
the flavon fields (which eventually determine
the mass matrix of $S$) can be expressed in the following matrix forms
in the basis $(S_1, S_2, S_3)^T$:
\begin{equation}
\sum_{k}y_{ijk}^{\phi}\phi_{k} =
y^{\phi}\left(\begin{array}{ccc}
0 & \phi_{2}-\phi_{3} & -\phi_{2}-\phi_{3}\\
\phi_{2}-\phi_{3} & \sqrt{2}\phi_{1} & 0\\
-\phi_{2}-\phi_{3} & 0 & -\sqrt{2}\phi_{1}
\end{array}\right),
\label{eq:gut-6}
\end{equation}
and
\begin{equation}
\sum_{k}y_{ijk}^{\xi}\xi_{k}=y^{\xi}\left(\begin{array}{ccc}
-e^{\frac{i\pi}{6}}\xi_{1}-\frac{\xi_{2}}{\sqrt{3}} & 0 & 0\\
0 & \frac{1}{6}\left(3e^{\frac{i\pi}{6}}\xi_{1}+\sqrt{3}\xi_{2}\right) &
\frac{1}{2}\left(e^{\frac{i\pi}{6}}\xi_{1}-\sqrt{3}\xi_{2}\right)\\
0 & \frac{1}{2}\left(e^{\frac{i\pi}{6}}\xi_{1}-\sqrt{3}\xi_{2}\right) &
\frac{1}{6}\left(3e^{\frac{i\pi}{6}}\xi_{1}+\sqrt{3}\xi_{2}\right)
\end{array}\right).
\label{eq:gut-6-1}
\end{equation}

To obtain non-trivial flavor structures, $S_{4}$ should be broken
down to  $(\mathbb{Z}_{2}\times\mathbb{Z}_{2})^{H}$ which differs
from $(\mathbb{Z}_{2}\times\mathbb{Z}_{2})^{V}$
$(\mathbb{Z}_{2}\times\mathbb{Z}_{2})^{H}$ is represented
by the matrices $R$ Eq.~(\ref{eq:gut-81-1}) in $\mathbf{3}$ of $S_{4}$ and
by $R^{(\mathbf{1})}$, $R^{(\mathbf{1}')}$, $R^{(\mathbf{2}')}$,
and $R^{(\mathbf{3}')}$
in the representations $\mathbf{1}$, $\mathbf{1}'$,$\mathbf{2}$,
and $\mathbf{3}'$, $(\mathbb{Z}_{2}\times\mathbb{Z}_{2})^{H}$
(see the appendix).
%%($R$ can also be denoted as
%%$R^{(\mathbf{3})}$. But for simplicity and consistency with the previous
%%discussion, we neglect the superscript $(\mathbf{3})$ if it does
%%not cause confusion).
Since the flavons $\phi$ and $\xi$ are assigned
to $\mathbf{3}'$ and $\mathbf{2}$ and break $S_{4}$ down
to $(\mathbb{Z}_{2}\times\mathbb{Z}_{2})^{H}$,
their vacuum expectation values (VEVs) should
be invariant under $(\mathbb{Z}_{2}\times\mathbb{Z}_{2})^{H}$, {\it i.e.}
\begin{equation}
R^{(\mathbf{3}')}\langle\phi\rangle\ = \langle\phi\rangle, ~~~~
R^{(\mathbf{2})}\langle\xi\rangle =  \langle\xi\rangle.
\label{eq:gut-112}
\end{equation}
This gives
\begin{equation}
\langle\phi\rangle\propto(0,\thinspace0,\thinspace1)^{T},\
\langle\xi\rangle\propto(0,\thinspace1)^{T},
\label{eq:gut-11}
\end{equation}
where we used explicit forms of $R^{(\mathbf{3}')}$ and $R^{(\mathbf{2})}$
from (\ref{eq:gut-17}) (\ref{eq:gut-18}).
%%the VEVs of $\phi$ and $\xi$ should be
%%to achieve the required symmetry breaking
%%$S_{4}\rightarrow$$(\mathbb{Z}_{2}\times\mathbb{Z}_{2})^{H}$.
The potentials which produce the vacuum alignment (\ref{eq:gut-11})
can be easily constructed \cite{Altarelli:2009gn}.
%\xj{to be added later}.
Finally from Eqs.~(\ref{eq:gut-6}), (\ref{eq:gut-6-1}) and (\ref{eq:gut-11}),
we obtain the explicit form of $M_{S}$, and consequently, $m_{\nu}$:
\begin{equation}
m_{\nu} \propto  M_{S} =
\left(\begin{array}{ccc}
a-2c & b & b\\
b & a+c & -3c\\
b & -3c & a+c
\end{array}\right),
\label{eq:gut-106}
\end{equation}
where
\begin{equation}
a=y^{\eta}\langle\eta\rangle,\ b=y^{\phi}\langle\phi\rangle,\
c=\frac{y^{\xi}}{2}\langle\xi\rangle.
\label{eq:gut-107}
\end{equation}
%that is $a$, $b$ and $c$
%come from contributions of $\langle\eta\rangle$,
%$\langle\phi\rangle$ and $\langle\xi\rangle$ respectively.

The mass matrix in Eq.~(\ref{eq:gut-106}) is diagonalized by $U_0 = U_{\rm{BM}}$ with
the eigenvalues
\begin{equation}
U_{0}^{T}m_{\nu}U_{0}=\left(\begin{array}{ccc}
a-\sqrt{2}b-2c & 0 & 0\\
0 & a+\sqrt{2}b-2c & 0\\
0 & 0 & a+4c
\end{array}\right).
\label{eq:gut-108}
\end{equation}
%%where $U_{0}$ is the BM mixing matrix given in Tab.~\ref{tab:rational}.
The three parameters $a$, $b$, and $c$ are sufficient to fit three light neutrino masses.
%Note that the scalar field $\eta$ is not necessary in this model.
%If $a$ in Eq.~(\ref{eq:gut-106}) is zero, there are only two
%parameters to fit three neutrino masses, which leads to a relation between
%the lightest neutrino masses:.. \xj{to be added later}.

%%%%%%%%%%%%%%%%%%%%%%%%%%%%%%%%%%%%%%%%%%%%%%%%%%%%%%%%%%%%%%%%%%%
\section{CKM mixing and PMNS mixings \label{sec:CKM}}
%%%%%%%%%%%%%%%%%%%%%%%%%%%%%%%%%%%%%%%%%%%%%%%%%%%%%%%%%%%%%%%%%%%%%%%%%%%

%%Our framework allows to disentangle  generation of  $U_0$ and the CKM mixing and some new physics should be
%%introduced to generate $V_{CKM}$.
Generation of $U_0$ and screening
in our framework require that the CKM-type  mixing originates from the  down components of the EW doublets.
Therefore, to reproduce the relation (\ref{eq:gut-45}), the mixing of the charged leptons
should be approximately equal to the down  quark  mixing:
$U_{l} \approx V_{\rm CKM}$.\footnote{We take the following convention
in the definitions of $U_{u}$, $U_{d}$,
$U_{e}$ and $U_{0}$: $M_{u}=U_{u}{\rm diag}(m_{u},\thinspace m_{c},\thinspace m_{t})U_{u}^{T}$,
$M_{d}=U_{d}{\rm diag}(m_{d},\thinspace m_{s},\thinspace m_{b})U_{d}^{T}$,
$M_{e}=U_{e}{\rm diag}(m_{e},\thinspace m_{\mu},\thinspace m_{\tau})U_{e}^{T}$
and $m_{\nu}=U_{0}^{*}{\rm diag}(m_{1},\thinspace m_{2},\thinspace m_{3})U_{0}^{\dagger}$.
In this convention, the CKM and PMNS matrices should be $U_{{\rm CKM}}
= U_{u}^{\dagger}U_{d}$, and $U_{{\rm PMNS}}=U_{l}^{T}U_{0}$.}
%%down quarks and charged leptons.
This   approximate equality of mixings should be reconciled with the difference of masses of down quarks and
charged leptons in the second and the first generations.
%%%%%%%%%%%%%%%%%%%%%%%%%%%%%%%%%%%%%%%%%%%%%%%%%%%%%%%%%%%%%%%%%%%%%%%%%%%%%%
%\begin{table*}
%\caption{\label{tab:masses} Fermion masses at the GUT scale ($2\times10^{16}$
%GeV) \cite{Das:2000uk}.}
%
%\begin{ruledtabular}
%\begin{tabular}{cccc}
%Notation & 1st generation & 2nd generation & 3rd generation
%\tabularnewline
%\hline
%$(m_{u},\thinspace m_{c},\thinspace m_{t})$ & $0.84_{-0.17}^{+0.16}$ MeV &
%$242.6_{-24.7}^{+23.6}$ MeV & $75.4_{-8.54}^{+9.96}$ GeV\tabularnewline
%$(m_{d},\thinspace m_{s},\thinspace m_{b})$ & $1.74_{-0.26}^{+0.48}$ MeV &
%$34.6_{-5.2}^{+4.9}$ MeV & $0.957_{-0.0169}^{+0.0037}$ GeV\tabularnewline
%$(m_{e},\thinspace m_{\mu},\thinspace m_{\tau})$ & $0.4413{}_{-0.0001}^{+0.0001}$ MeV
%& $93.143{}_{-0.010}^{+0.014}$ MeV & $1.5834{}_{-0.0005}^{+0.0001}$ GeV
%\tabularnewline
%\end{tabular}
%\end{ruledtabular}
%\end{table*}
%%%%%%%%%%%%%%%%%%%%%%%%%%%%%%%%%%%%%%%%%%%%%%%%%%%%%%%%%%%%%%%%%%%%%%%%%%%%
%%
In fact, according to the two-loop  RGE running in the Standard Model \cite{Das:2000uk}
we have at the GUT scale
\begin{equation}
m_{\mu}\approx3m_{s}\gg m_{e},\thinspace m_{d}.
\label{eq:gut-34}
\end{equation}
This problem was extensively discussed before in connection
to the  quark-lepton complementarity \cite{Antusch:2005ca}.
Actually, in the case of strong mass hierarchy the difference of mixings related
to the difference of masses is not large and
may be even needed to better fit of the data.

%%Equal mixings of d-quarks and charged leptons can be achieved if the unique Higgs multiplet (or several
%%multiplets in the same representation) produce dominant (unique) contribution to mass matrices.
%%In this case, however,  the masses of quarks and leptons from the same $\psi$
%%turn out to be proportional:  $(m_{e}:m_{\mu}:m_{\tau})=(m_{d}:m_{s}:m_{b})$.

%\note{Antusch}\xj{not found: Antusch's paper with QLC}.

In general,  there are two approaches to keep the relation $U_{l} \approx  U_{d}$ for different masses.
The one is to decouple completely the generation of masses from mixing,
so that the mixing comes from certain relations between
the elements of mass matrix,  whereas the masses
are determined by absolute values of the elements.
This decoupling is difficult to obtain for small CKM mixing:
simple discrete symmetries usually lead to large nonzero angles\footnote{There are, however,
some finite groups contain small angles---see
e.g. \cite{Toorop:2011jn,deAdelhartToorop:2011re,Rodejohann:2015pea}.}.
Hence this approach would require substantial complications of the model.
Another possibility, which we will implement here,
is that one of the Higgs multiplets dominates in
the generation of charged fermion masses, so that $M_{d}$ and $M_{l}$
have roughly the same form, and thus approximately equal mixing.
One can  also add  a mass matrix proportional to the unit matrix: this does not change the 
mixing but affects the mass ratio.

%%In what follows we will present the simplest scenario and mark other possibilities.

With  one 10-plet, which conserves the basis fixing symmetry, we obtain at the GUT scale the diagonal mass matrices:
%%
%%the large masses of the third generation to the renormalizable operator
%%$\psi\psi H^{(10)}$ in the presence of a fundamental 10-plet Higgs
%%that only generates the following mass matrix
\begin{equation}
M{}_{d}^{(10)} = M{}_{l}^{(10)} =  \frac{v_d}{v_u} M{}_{u}^{(10)} =
\left(\begin{array}{ccc}
k_1 & 0 & 0\\
0 & k_2 & 0\\
0 & 0 & k_3
\end{array}\right),
\label{eq:gut-51}
\end{equation}
where $v_d$ and $v_u$ are the VEVs of the 10-plet which generate masses of the upper and bottom  components of the EW doublets correspondingly.
The difference of masses of up and down components is due to the difference of VEVs:  $v_d/v_u \approx m_b/m_t$.
Since $k_1 : k_2 : k_3 = m_u : m_c : m_t$,  we obtain
for  $k_3 \sim m_b \sim m_\tau \sim 1$ GeV, and 
%%fixed by the VEV of the down-type  Higgs doublet from the 10-plet.
%%For other mass terms we find
the other masses $k_2 \approx 3$  MeV,  $k_1 \approx 10^{-2}$ MeV,
which are much smaller than $m_s$ and $m_d$ correspondingly.
So,  additional sources of mass and mixing are needed. The simplest possibility is
that the total matrices of the down-type quarks and charged leptons consist of
\begin{equation}
M_{d} = M_{d}^{(10)}+M_x,\ ~~~~~ M_{l}= M_{d}^{(10)}+ a M_x,
\label{eq:gut-32}
\end{equation}
where  $|a| \approx 3$ is needed to reproduce  (\ref{eq:gut-34}),  and maximal values of elements in $M_x$
should be $M_x^{max} \sim m_s$ not to destroy the $b$-$\tau$ unification.
The matrix $M_x$ is non-diagonal, thus breaking the basis fixing symmetry or $G_V$, and producing  the CKM mixing.

Correct masses and mixing can be obtained provided that  $M_x$ has the structure
\begin{equation}
M_x \approx \left(\begin{array}{ccc}
d_{1}   &  f  & f'\\
f  & d_{2} &  d'\\
f' & d' & d_{3}
\end{array}\right),
\label{eq:126cont}
\end{equation}
with
\begin{equation}
f' \approx f, ~~~~ d' \sim d_2 \sim d_3,~~~~~ d_1 \ll d_2,  d_3,
\label{eq:126elem}
\end{equation}
and
\begin{equation}
f  \approx d_2 \sin \theta_C.
\label{eq:126elem}
\end{equation}
That is, the mass matrix $M_x$ (\ref{eq:126cont}) has the  dominant 2-3 block and the Cabibbo suppressed
12-and 13-elements. It is  similar to the TBM or BM  mass matrices.
Therefore it  is also similar to the structure of $M_S$, and consequently, to $m_\nu$,
in the case of normal mass hierarchy.
It is interesting to speculate that common  Planck scale physics is responsible for the
structure of $M_S$ and $M_x$.

Numerically we need to have
\begin{equation}
d_i  \sim 0.1 v_{EW} \frac{M_{GUT}}{M_{Pl}} \sim (30 - 100) ~{\rm MeV},
\label{eq:126ddd}
\end{equation}
comparable to the masses of  second generation, {\it i.e.} muon and $s$-quark.

The total mass matrices  of the down quarks
and charged leptons (\ref{eq:gut-32}) for $a = -3$ become
\begin{equation}
M_{d} \approx \left(\begin{array}{ccc}
d_{1}+k_{1} & f  & f'\\
f  & d_{2}+k_{2} & d'\\
f' & d' & d_{3}+k_{3}
\end{array}\right),\
M_{l} \approx \left(\begin{array}{ccc}
-3d_{1}+k_{1} & -3f & - 3f'\\
-3f & -3d_{2}+k_{2} & - 3d'\\
- 3f' & - 3d' & -3d_{3}+k_{3}
\end{array}\right),
\label{eq:gut-41}
\end{equation}
where $d_{1} \gg  k_{1}$ and  $d_{2} \gg k_{2}$.
So, for the second and the first
generations the contributions from $M_x$  dominate.
%%The mass terms $d_i$ and  $f$ that appear with the factor 3 in the lepton matrix,
%%are due to the effective  126-plet,  whereas the diagonal elements
%%$k_i = h_i + k_i^{eff}$ are contributions from 10-plets (fundamental and effective).
This leads to (i) about three times larger mass of muon than
the mass of $s$ quark,  and (ii) to approximately the same 1-2  mixing
of leptons and quarks.
For the third generation, the contribution from the
10-plet dominates, $k_3 \approx h_3 \gg d_3$, thus ensuring the approximate $b$\,-\,$\tau$ unification.
In general,  $m_{\tau} = m_{b}+ {\cal O}(4 m_{\mu})$.
%instead of Eq.~(\ref{eq:gut-53}).

From (\ref{eq:gut-41}) we obtain for the 2-3 quark mixing
\begin{equation}
V_{cb} \approx \frac{d'}{k_3}  \approx  \frac{d'}{m_b} \approx \frac{m_s}{m_b},
\label{eq:q23}
\end{equation}
and  the 1-3 quark mixing
\begin{equation}
V_{ub} \approx \frac{f'}{k_3}  \sim  \frac{f}{m_b} \sim \frac{m_s}{m_b} \sin \theta_C,
\label{eq:q12}
\end{equation}
in agreement with observations.
According to (\ref{eq:q23})  and (\ref{eq:q12}),  $V_{ub} \sim V_{cb} V_{us}$.

The lepton mixing parameters are  about 3 times larger:
\begin{equation}
U_{\mu 3} \approx -  \frac{3d'}{m_\tau},  ~~~~
U_{e 3} \approx -   \frac{3f'}{m_\tau}.
\end{equation}
This corresponds to the angles  $\theta_{23}^l \sim (4 - 5)^\circ$ and $\theta_{13}^l \sim 1^\circ$,
which give a sizable deviation from maximal 2-3 mixing and
observable corrections to the 1-3 mixing.

Decoupling of the third state produces small  corrections to the 1-2 sub-matrix of $M_d$:
The correction to the 1-1 element $f'^2/m_b \sim 0.05$ MeV,
the relative corrections to other  elements are  of the order $d'/m_b \sim m_s/m_b \sim 3\%$
and can be neglected.
The corresponding relative corrections in the lepton sector are 3 times larger;
the correction to the 1-1 element which is an order of magnitude larger:  $9 f'^2/m_b \sim 0.45$ MeV can be important  for the mass of electron.
In the 1-2 sector we can reproduce the Gatto-Sartori-Tonin relation:
$\sin \theta_C \sim \sqrt{m_d/m_s}$.

Let us make few comments on possible origins of  $M_x$.
The straightforward way is to introduce a 126-plet which produces $a = - 3$ in Eq. (\ref{eq:gut-32}).
This  126-plet  should  not contribute
substantially to the masses of neutrinos,  not to destroy the inverse seesaw with
screening. For this, the VEVs of the $SU(2)$ singlet and triplet in 126-plet should be zero or small.
The mass of 126-plet can be at Planck scale to avoid the  problem of perturbativity
of the theory  (see, e.g., the review \cite{Senjanovic:2005sf}).

Another possibility \cite{Babu:1998wi}
is to use composite 126-plet constructed from the product of %$\mathbf{16}\times\mathbf{16}$ Higgs multiplets
two 16-plets. The coupling with fermions is given by
non-renormalizable operators suppressed by
%some scale, which we assume to be
the Planck scale $M_{{\rm Pl}}$:
\begin{equation}
{\cal L}\supset  \frac{1}{M_{\rm Pl}}  \psi \psi H^{(16)}H'^{(16)}.
\label{eq:gut-48}
\end{equation}
Here $H'^{(16)}$ is new 16-plet of scalars with zero VEV of the $SU(2)$ triplet and singlet components.
Similar operator  with $H^{(16)}H^{(16)}$ can be forbidden by additional symmetry
with respect to transformations $H^{(16)}  \rightarrow i H^{(16)}$,
$H'^{(16)}  \rightarrow - i H'^{(16)}$, $S \rightarrow - iS$, $(\eta,\, \xi,\, \phi) \rightarrow - (\eta,\,  \xi,\, \phi)$.
Then one should assume that due to some Planck scale physics the down Higgs doublet in the composite
126-plet  acquire the VEV
\begin{equation}
\langle [ H^{(16)} H^{(16)'}]^{(126)} \rangle{}_d = v_d {M_{GUT}}.
\label{eq:vevdhiggs}
\end{equation}
For $v_d \sim 0.1 v_{EW}$, this reproduces Eq.~(\ref{eq:126ddd}).
%%%%%%%%%%%%%%%%%%%%%%%%%%%%%%%%%%%%%%%%%%%%%%%%%%%%%%%%%%%%%%%%%%%%%%%%%

Notice that  instead of the non-renormalizable interaction (\ref{eq:gut-48}) we can introduce
\begin{equation}
{\cal L}\supset  \frac{1}{M_{\rm Pl}}  \psi \psi H^{(10)}H^{(45)},
\label{eq:gut-48aa}
\end{equation}
where $H^{(45)}$ is the 45-plet responsible for the SO(10) symmetry  breaking  \cite{Babu:1998wi}.
The product  $H^{(10)}H^{(45)}$ contains antisymmetric 120-plet, and therefore can remove the degeneracy of 
charge lepton and d-quark masses. However,
the matrix (\ref{eq:126cont}) with diagonal elements
can not be  reproduced.  \\

%%%%%%%%%%%%%%%%%%%%%%
In what follows, for simplicity we will consider  mixing of the first two generations only.
Inclusion of corrections from the 1-3 or 2-3 mixing changes the following results
very little. The mass matrices (\ref{eq:gut-41}) can be diagonalized by
%xxj: U or V
\begin{equation}
U_{d}=\left(\begin{array}{ccc}
c & \tilde{s} & 0\\
-\tilde{s}^{*} & c & 0\\
0 & 0 & 1
\end{array}\right)P_{d},\ ~~~~ U_{l}=\left(\begin{array}{ccc}
c_{l} & \tilde{s}_{l} & 0\\
-\tilde{s}_{l}^{*} & c_{l} & 0\\
0 & 0 & 1
\end{array}\right)P_{l},\
\label{eq:gut-36}
\end{equation}
where
$$
c \equiv\cos\theta_{C},\ \
\tilde{s} \equiv\sin\theta_{C}e^{i\phi_{C}}, \ \
c_{l}\equiv\cos\theta_{l},\ \
\tilde{s}_{l}=\sin\theta_{l} e^{i\phi_{l}},
$$
and $P_{d}$ and $P_{l}$
are diagonal matrices containing complex phases.
Although all elements of the mass matrices in Eq.~(\ref{eq:gut-41})
are complex, for simplicity, we assume that only $f$ is complex.
Then six real parameters $d_{1,2,3}$ and $k_{1,2,3}$ allow us
to accommodate the six masses
$(m_{d},\thinspace m_{s},\thinspace m_{b},\thinspace m_{e},\thinspace m_{\mu},m_{\tau})$,
while the complex  $f$ generates the Cabibbo mixing, $\sin \theta_{C}$,
with a complex phase $\phi_{C}$, and analogous mixing $\sin \theta_{l}$ and phase $\phi_{l}$ in the lepton
sector.

The phase $\phi_{C}$ has no physical meaning for the
$2 \times 2$ form of $U_{d}$.  In contrast, as we will see, $\phi_{l}$
is directly related to the CP phase in the PMNS matrix.
Introduction of small 1-3 and 2-3 mixing will make $\phi_{C}$ to be
the origin of CP violation in the CKM mixing. But this will have little
effect on the PMNS mixing.

Using  the hierarchy $m_{d}\ll m_{s}$ and $m_{e}\ll m_{\mu}$ as
well as the smallness of  $\sin\theta_{C}\ll1$, we obtain the following
approximate relations (for more details, see  Appendix \ref{sec:analytic}):
\begin{equation}
\frac{\phi_{l}}{\phi_{C}} = 1 + {\cal O}\left(\frac{m_{d}}{m_{s}}\right),
\label{eq:gut-65}
\end{equation}
\begin{equation}
\frac{\sin 2\theta_{l}}{\sin 2\theta_{C}}\approx\frac{3\left(m_{s} +
m_{d}\cos\phi_{1}\right)}{m_{\mu}+m_{e}\cos\phi_{2}}\approx1,
\label{eq:gut-66}
\end{equation}
where
\begin{equation}
\phi_{1}\equiv\pi-\arcsin\left(\frac{s^{2}m_{s}}{c^{2}m_{d}}\sin2\phi_{l}\right)-2\phi_{l},\
~~\phi_{2}\equiv\arcsin\left(\frac{s_{l}^{2}m_{\mu}}{c_{l}^{2}m_{e}}\sin2\phi_{l}\right)-2\phi_{l}.
\label{eq:phi12}
\end{equation}
Eq.~(\ref{eq:phi12}) shows complicated dependence of the phases, on  known quantities
(fermion masses, the Cabibbo angle) and on $\phi_l$ which in turn is related to the leptonic CP phase.
Values of $\phi_{1}$ and $\phi_{2}$ for two special values of $\phi_{l}$
can be obtained from (\ref{eq:phi12}),
\begin{equation}
(\phi_{1},\ \phi_{2})\approx\begin{cases}
(0^{\circ},\ 180^{\circ}) & \ {\rm for}\ \phi_{l}=\pm90^{\circ}
~~~~~~~~~({\rm i})\\
(180^{\circ},\ 0^{\circ}) & \ {\rm for}\ \phi_{l} =
0^{\circ}\thinspace{\rm or}\thinspace180^{\circ}
~~~({\rm ii})
\end{cases}.
\label{eq:gut-76}
\end{equation}
For other values of $\phi_{l}$ results of
numerical study will be presented later.

Eqs.~(\ref{eq:gut-65}) and (\ref{eq:gut-66}) show that the charged
leptons do have approximately the same mixing as the down-type quarks,
$\theta_{l}\approx\theta_{C}$ and $\phi_{l}\approx\phi_{C}$.
Recall that the factor $3$ in Eq.~(\ref{eq:gut-66}) originates from
the effective 126-plets.
Taking the 1$\sigma$ range values of $m_{e}$, $m_{\mu}$,
$m_{u}$, $m_{s}$ from
Ref.~\cite{Das:2000uk}, and using Eq.~(\ref{eq:gut-66}) we can evaluate
the  ratio of the angles for the two choices of phases [cf.  Eq.~(\ref{eq:gut-76})]:
\begin{equation}
\frac{\theta_{l}}{\theta_{C}} =
\begin{cases}
0.871  - 1.22\,\  ({\rm i})\\
0.999 - 1.35 \, \   ({\rm ii}).
\end{cases}
\label{eq:gut-76-x}
\end{equation}
For other values of the phases, one would get intermediate results
between those in the cases (i) and (ii).

%%%%%%%%%%%%%%%%%%%%%%%%%ffff1%%%%%%%%%%%%%%%%%%%%%%%%%%%%%%%%%%%%%%%%%%%%%%%%%%%%%55
\begin{figure}
\centering

\includegraphics[width=8cm]{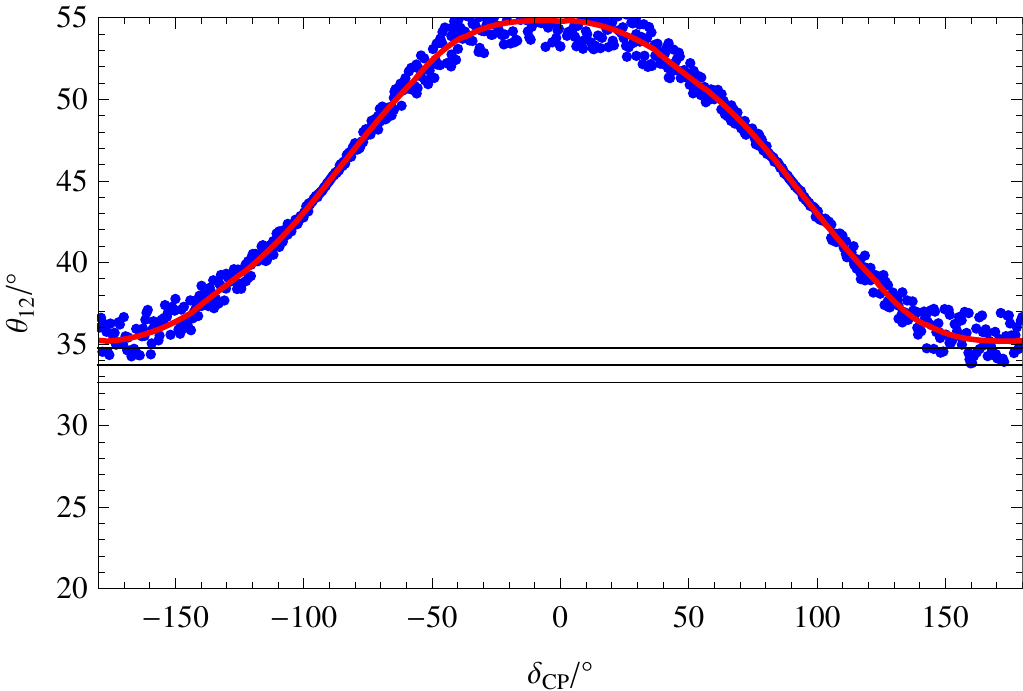}~~~ \includegraphics[width=8cm]{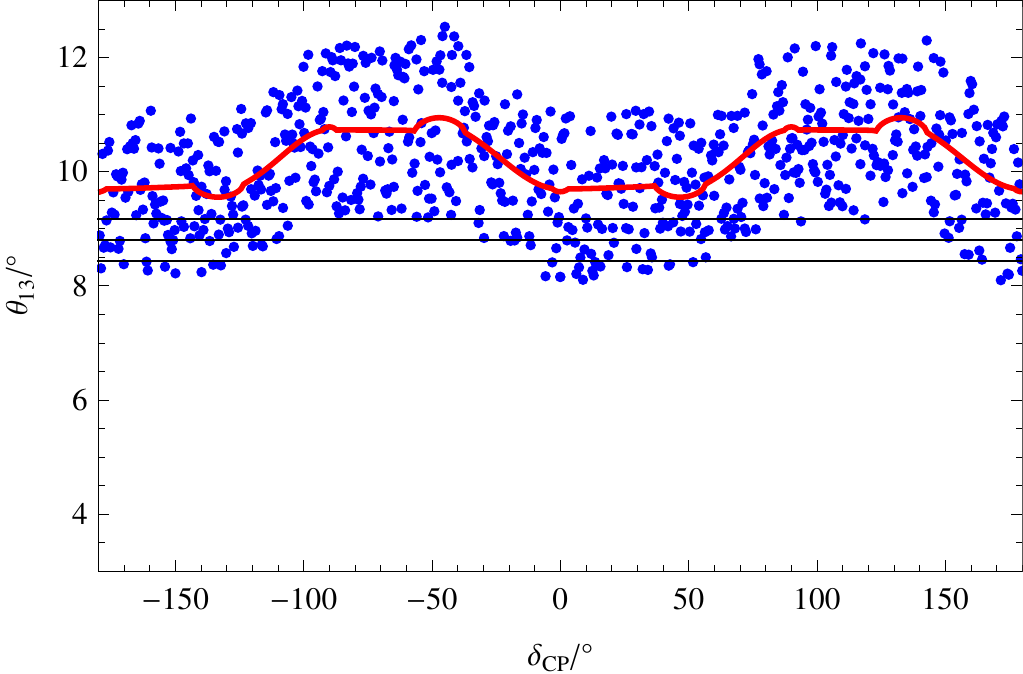}

\caption{\label{fig:Predictions}Predictions on the PMNS mixing angles.
Shown are the dependencies of the angles $\theta_{12}$ (left panel) and $\theta_{13}$
(right panel) on $\delta_{CP}$.
Red lines correspond to the
charged fermion masses fixed at the central values according to the RGE running in \cite{Das:2000uk};
blue points are  masses randomly generated within 1$\sigma$ allowed regions.
The black lines show the  best-fit values and 1$\sigma$ bounds of $\theta_{13}$ and $\theta_{12}$
from \cite{Patrignani:2016xqp}. We take $\sin \theta_C = 0.225$. }
\end{figure}
%%%%%%%%%%%%%%%%%%%%%%%%%%%%%%%%%%%%%%%%%%%%%%%%%%%%%%%%%%%%%%%%%%%%%%%%%%%%%%

%For the CKM contribution we only consider the 1-2 mixing.
According to Eq.~(\ref{eq:gut-36}),  the PMNS matrix
should be
\begin{equation}
U_{{\rm PMNS}}=\left(\begin{array}{ccc}
c_{l} & -\tilde{s}_{l}^{*} & 0\\
\tilde{s}_{l} & c_{l} & 0\\
0 & 0 & 1
\end{array}\right)  U_{{\rm BM}}
\label{eq:gut-115}
\end{equation}
%where $\tilde{s}_{l} = \sin \theta_l e^{i\phi_l}$
%is related to the CKM mixing by Eqs.~(\ref{eq:gut-65})
%and (\ref{eq:gut-66}).
%From Eq.~(\ref{eq:gut-115}), we get
or explicitly
\begin{equation}
U_{{\rm PMNS}}=\left(\begin{array}{ccc}
\frac{1}{2}\left(\tilde{s}_{l}^{*}+\sqrt{2}c_{l}\right) &
-\frac{\tilde{s}_{l}^{*}}{2}+\frac{c_{l}}{\sqrt{2}} & \frac{\tilde{s}_{l}^{*}}{\sqrt{2}}\\
\frac{\tilde{s}_{l}}{\sqrt{2}}-\frac{c_{l}}{2} & \frac{1}{2}\left(\sqrt{2}\tilde{s}_{l}
+ c_{l}\right) & -\frac{c_{l}}{\sqrt{2}}\\
-\frac{1}{2} & \frac{1}{2} & \frac{1}{\sqrt{2}}
\end{array}\right).
\label{eq:pmns-f}
\end{equation}
Notice that the matrix of phases $P_l$ does not appear here since it can
be removed by rephasing of the fields.
From (\ref{eq:pmns-f}) one finds mixing parameters in the standard parametrization,
\begin{equation}
s_{13}=\frac{s_{l}}{\sqrt{2}},
\label{eq:gut-116}
\end{equation}
\begin{equation}
s_{12}^2 = \frac{1}{2}-\frac{\sqrt{2}c_{l}s_{l}\cos\phi_{l}}{2-s_{l}^{2}},
\label{eq:gut-117}
\end{equation}
\begin{equation}
s_{23}^2 = \frac{c_l^2}{2-s_{l}^{2}} \approx
\frac{1}{2}\left( 1 - \frac{1}{2} s_l^2\right),
\label{eq:gut-23a}
\end{equation}
\begin{equation}
\sin\delta_{{\rm CP}}=-\sin\phi_{l}-s_{l}^{2}\sin\phi_{l}\cos^{2}\phi_{l} +
{\cal O}(s_{l}^{3}).
\label{eq:gut-118}
\end{equation}
The above results can be expressed in terms of known variables ($\theta_{C}$,
$m_{s}$, $m_{d}$, $m_{\mu}$, $m_{e}$) and $\phi_{l}$. Although
$\phi_{l}\approx\phi_{C}$, we cannot connect it to the CP violation
in the CKM mixing without introducing 1-3 and 2-3 mixing.
Using expression Eq.~(\ref{eq:gut-73}) for the mixing parameter $s_l$
and replacing $\phi_{l}$ by $\delta_{{\rm CP}}$, we obtain from (\ref{eq:gut-116})
\begin{equation}
s_{13}\approx3\frac{\sin\theta_{C}}{\sqrt{2}}
\left|\frac{m_{s}-m_{d}e^{-2i\delta_{{\rm CP}}}e^{-i\sin^{-1}
\left(\frac{m_{s}}{m_{d}}\tan^{2}\theta_{C}\sin2\delta_{{\rm CP}}\right)}}{m_{\mu} +
m_{e}e^{-2i\delta_{{\rm CP}}}e^{i\sin^{-1}
\left(\frac{m_{\mu}}{m_{e}}\tan^{2}\theta_{C}\sin2\delta_{{\rm CP}}\right)}}\right|.
\label{eq:gut-119}
\end{equation}
From (\ref{eq:gut-116})  and (\ref{eq:gut-117}) we obtain the relation between observables:
\begin{equation}
s_{12}^2 \approx \frac{1}{2}+\frac{s_{13}\cos\delta_{{\rm CP}}}{c_{13}^{2}}.
\label{eq:gut-120}
\end{equation}
In Fig.~\ref{fig:Predictions} we show the mixing angles
$\theta_{12}$ and $\theta_{13}$ as functions of
the CP phase $\delta_{\rm CP} \approx \phi_{l}$  according
to Eqs.~(\ref{eq:gut-117}) and  (\ref{eq:gut-116}).
We used $\sin \theta_{C}=0.225$ and
the values of $m_{s}$, $m_{d}$, $m_{\mu}$, $m_{e}$ from Ref.~\cite{Das:2000uk}.
%Tab.~\ref{tab:masses},
The angle $\theta_{23}$ is in the first octant: $\sin^2 \theta_{23} =  0.49$.
Taking $1\sigma$ allowed interval for  $\theta_{12}$
we obtain from (\ref{eq:gut-120})  $\cos \delta_{CP} < - 0.86$  or
\begin{equation}
%%144^{\circ}\lesssim\delta_{\rm CP}\lesssim 210^{\circ}.
\delta_{CP} \in (0.80\pi, 1.16\pi).
\label{eq:cpphase}
\end{equation}
Including 2-3 mixing in $V_{\rm CKM} \simeq U_{l}$ changes the prediction
for $\delta_{\rm CP}$ by a few degrees.
The result (\ref{eq:cpphase}) is in agreement with general phenomenological
analysis  \cite{Petcov:2017ggy} for the case of BM mixing receiving corrections
from charge lepton mixing  $U_l$.
%%The result in Ref. \cite{Petcov:2017ggy} shows
%%that for general 1-2 and 2-3 mixing considered in $U_l$,
%%to obtain correct values of $(\theta_{12}, \theta_{13}, \theta_{23})$,
According to  \cite{Petcov:2017ggy} $\cos\delta_{\rm CP}$
should be in the range $[-1.00,\,-0.72]$ at $3\sigma$ confidence level.

The upper value of the interval  (\ref{eq:cpphase}) is in agreement with global
fit results at about $1 \sigma$ level.
Notice that  $\delta_{CP}$ is  strongly restricted here
by the requirement of  maximal possible reduction of the 1-2 mixing
from its BM value $\sin^2 \theta_{12} = 0.5$.
So, that the best value would be $\cos \delta_{CP} = -1$.
The only other parameter that enters the relation
(\ref{eq:gut-120}) is the 1-3 mixing which is measured very precisely.
If we would use the TBM matrix $U_0 = U_{TBM}$
instead of BM, no large corrections from $U_l^\dagger$
is required and $\cos \delta_{CP}$ should be close to zero.
However, $U_0 = U_{TBM}$ can not be obtained in our
residual symmetry approach.
Thus, future measurements of $\delta_{CP}$
will test the scenario.

The values of  mixing angles are subjects of the renormalization group (RG) corrections.
The CKM mixing receives small corrections. For the PMNS mixing
 in our framework the RG corrections can be significant
due to strong hierarchy of the  right-handed neutrino masses.
The strong hierarchy originates  from the up-type quark mass spectrum,
as the right-handed neutrino mass matrix is determined by $m_{D}^{T}m_{\nu}^{-1}m_{D}$
and  in $SO(10)$ models, $m_{D}$ is the same as $M_{u}$ (or approximately
the same in multi-Higgs variations). From this one obtains
the masses of the order  %$(1,\thinspace10^{-5},\thinspace10^{-10})\times10^{14}$
$(10^4,\thinspace10^{9},\thinspace10^{14})$
GeV.  Between the lightest
and heaviest right-handed neutrinos, one or two of them are integrated
out while the others remain in the RG equations. This is where RG
running may have larger effects on the flavor structure \cite{Antusch:2005gp}.

The RG effects with such a strong
hierarchy have been studied in \cite{Schmidt:2006rb}. In the SM extended by RH neutrinos,
the corrections mainly depend on the lightest neutrino mass, Majorana phases and
the mass ordering.  The result in \cite{Schmidt:2006rb}
shows that for $m_{1}=10^{-3}{\rm \ eV}$ in the case of normal mass ordering,
the correction to $\theta_{12}$ is  in the range $(-1.5^{\circ},\thinspace1.0^{\circ})$.
For smaller/larger $m_{1}$, the correction can be significantly
suppressed/enhanced. For example, when $m_{1}=10^{-4}{\rm \ eV}$, the correction
can be reduced down to $0.02^{\circ}$ (cf. Fig.~10 in \cite{Schmidt:2006rb}),
independent of the Majorana phases; while for $m_{1}=10^{-2}{\rm \ eV}$
with zero Majorana phases, it can reach $10^{\circ}$.

The other two mixing angles $\theta_{13}$ and $\theta_{23}$ are
generally much more stable with respect to the RG corrections than $\theta_{12}$
\cite{Antusch:2003kp,Rodejohann:2015nva,Rodejohann:2017lre}.
In the scenario with strong hierarchy of right-handed neutrino masses,
the correction to the 1-3 mixing is found to be always smaller than $0.3^{\circ}$
in the SM,  which is negligible compared to the uncertainties caused by
the fermion masses at the GUT scale\,---\,see Eq.~(\ref{eq:gut-76-x}).

The fermion singlets $S$ from  the hidden sector may produce further
corrections. Since the mass scale of $S$ is not much higher than
the GUT scale, we expect small RG corrections from, e.g. $10^{16}$
GeV to $10^{18}$ GeV.
At the GUT scale or below, the singlets can
be integrated out, generating the heavy Majorana masses of right-handed
neutrinos.
Therefore, we can assume that the SM plus type I seesaw
is valid up to the GUT scale and in this range, one can adopt  the conclusions
from \cite{Schmidt:2006rb}. Above the GUT scale, up to the mass scale
of $S$ we expect small  RG corrections
due to small interval or running. Furthermore, the mixing of these singlets
with active neutrinos is strongly suppressed.

Variations of the CP phase predictions may be possible.
%%which is in some tension with  indicated maximal CP-violation.
Essentially the result (\ref{eq:cpphase})
is obtained in assumption of negligible RG corrections.
Large RG corrections $(\sim 10^{\circ})$ to the 1-2 mixing
can be obtained for  the degenerate pair of $\nu_1$ and $\nu_2$.
This can be realized for $m_1 \sim 10^{-2}$ eV or inverted mass ordering.
In this case RG corrections can reduce $\theta_{12}$ down to
$35^{\circ}$ and  large corrections from $U_l$ are not needed.
Consequently,  $\cos \delta_{CP}$ can be small.

One can use some other matrices from the Table I,
e.g. $U_{GR}$.
%%[[this may be different from notation in Petcovs papers]]
This however, will require large corrections from $U_l^{\dagger}$
and its substantial deviation from $V_{\rm CKM}$.
Another possibility is to produce some mixing from the portal interactions.
Finally one can abandon the residual symmetry approach
and fix structure of $M_S$ using other symmetries
or principles.

Finally let us  comment on the viability of leptogenesis in this framework.
%since the three RH neutrino masses have the strong hierarchy,
As previously mentioned, the typical RH neutrino masses
are $(10^4,\thinspace10^{9},\thinspace10^{14})$
GeV, which is very hierarchical.  The lightest RH neutrino in this scenario
is too light to produce the observed baryon asymmetry \cite{Davidson:2002qv}.
However, the second RH neutrino is in the proper mass range
and may realise successful   $N_2$-leptogenesis (see, e.g., \cite{DiBari:2015svd,DiBari:2015oca}).

%%%%%%%%%%%%%%%%%%%%%%%%%%%%%%%%%%%%%%%%%%%%%%%%%%%%%%%%%%%%%%%%%%%%%%%%%%%
\section{Conclusions \label{sec:Conclusion}}
%%%%%%%%%%%%%%%%%%%%%%%%%%%%%%%%%%%%%%%%%%%%%%%%%%%%%%%%%%%%%%%%%%%%%%%%%%%%%%%%

1. The relation between the lepton and quark mixings
[cf. Eq.~(\ref{eq:gut-45})] can imply the Grand Unification and existence of the hidden sector
which is connected to the visible sector via the RH neutrino portal.
The Grand Unification ensures the approximate equality
$U_l \approx  V_{d} \approx V_{\rm CKM}$, and consequently, $U_{\rm PMNS} \simeq V_{\rm CKM}^\dagger U_0$,
whereas the hidden sector with certain symmetries generates $U_0$ and
produces the smallness of neutrino masses.

2. We focus on the symmetry aspects of this scenario\,---\,the interplay of discrete
flavor symmetries and the SO(10) gauge symmetry.
We develop the residual symmetry approach to generate
$U_0$ which connects the  visible and hidden sectors.
The $\mathbb{Z}_{2}\times\mathbb{Z}_{2}$  residual symmetries of the visible and hidden sectors
are intrinsic symmetries of the SO(10) Yukawa interactions.

3. Embedding of these residual symmetries into a unified finite flavor group
fixes the moduli of mixing matrix elements in the form of cosines
of rational multiples of $\pi$. Imposing the unitarity condition
results in  only a few forms of $U_0$ which include
the BM  matrix.
%, the matrix determined by Golden ratio, etc.
Using these matrices we reconstructed the group presentations and
thus identified the corresponding symmetry groups.
The flavor symmetry  is broken in the hidden sector spontaneously.
In the visible sector (at lower energy scales) the breaking can be explicit.
In the latter case only
the basis symmetry in the hidden sector
is promoted to a larger non-abelian symmetry.
%%This most economical version of the breaking produces small corrections to masses and mixing.
%%Here we use the fact that the residual symmetries are realized
%%for the visible and hidden interactions at different energy scales.
We considered a specific model which realizes the BM mixing for $U_0$.

4. We use  the 126-plet with the Planck scale mass or composite  126-plet originating
from the Plank-scale physics
%%and interactions via  non-renormalizable operators $\psi \psi H^{(16)} H^{(16)'}$
to generate the CKM mixing
and $U_l \sim V_{CKM}$ as well as differences  of quark and lepton masses of the second and the first generations.
Thus, generation of CKM mixing and the mass differences of the down quarks
and charge leptons of the first the second
generations are connected.
Interestingly, the CKM mixing and the corresponding $U_{l}$ matrix can be reproduced
with additional contribution from the Planck-scale physics  with the flavor structure similar
to the one for $S$ and consequently, the  light neutrinos.
%%Both masses of  $S$ and $M^{(126)}$ originate from the Planck scale physics
%%and so can be indeed connected.

5. Assuming that the RGE corrections are small we expect
the leptonic CP violation phase to be in the range
$144^{\circ}\lesssim\delta_{\rm CP}\lesssim 210^{\circ}$.
Future measurements of $\delta_{\rm CP}$ in accelerator neutrino
experiments such as T2K, NOVA and DUNE will be an important test of this scenario.

6. Coupling of the hidden sector with visible
one realizes the double seesaw mechanism. It allows to disentangle generation  of the
$U_0$ mixing and the CKM mixing.
The latter is related to the Planck scale
suppressed non-renormalizable interactions.

7. An important feature of this scenario is a very strong hierarchy
of masses of the RH neutrinos. This can lead to significant
renormalization group effects which correct, in particular,
the 1-2 mixing.
The lightest RH neutrino with mass $10^4$ GeV has mixing
with active neutrinos of the order $10^{-7}$. So,  it can not be
observed at colliders, but could play some role in leptogenesis.
The $N_2$ leptogenesis can be realized.

8.  Future precise measurements of the CP phase,
establishing mass ordering and absolute scale
(degeneracy) of masses will provide important tests
of the scenario.  In particular, establishing
strong normal mass hierarchy and substantial CP violation
would exclude the simplest realization based on the BM mixing from the
hidden sector.

No new physics related to the neutrino mass generation should be
observed at LHC and other future collider experiments.
Proton decay might be detected at some level.

%
%5. We briefly outline generic features
%of the scenario and its possible tests.

%%%%%%%%%%%%%%%%%%%%%%%%%%%%%%%%%%%%%%%%%%%%%%%%%%%%%%%%%%%%%%%%%%%%%%%%%%%%%%%%
%%Two different new physics are involved in
%%generation of the relation (\ref{eq:gut-45}): one is the CKM new physics
%%acting in both quark and lepton sectors and another one is the neutrino
%%new physics which produces smallness of neutrino masses
%%and $U_0$ with certain symmetry.

%%Mixing is generated by the hidden sector.

%%This is the residual symmetry approach applied to the visible and hidden sectors.

%%We have found all mixing matrices with $p\leq100$, which satisfy
%%the symmetry group relations and unitarity. They include
%%the trivial identity matrix, infinite numbers of block diagonal
%%matrices, the BM mixing matrix, and three matrices in which the entries
%%can only be $\cos36^{\circ}$, $\cos45{}^{\circ}$, $\cos60^{\circ}$,
%%and $\cos72^{\circ}$.

%[[The main idea was to keep in the first approximation
%all the matrices to be diagonal apart from mass matrix for $S$ which gives
%$U_{\rm{PMNS}} = U_0$ in the first approximation.]]

%%$(G_V = \mathbb{Z}_{2}\times\mathbb{Z}_{2})^{v}$.

%%%%%%%%%%%%%%%%%%%%%%%%%%%%%%%%%%%%%%%%%%%%%%%%%%%%%%%%%%%%%%%%%%%%%%%%
\begin{acknowledgments}
XJX would like to thank Patrick Ludl who has done an early work in  this framework. The author appreciates many discussions with
him, both on physics and industry.
\end{acknowledgments}

\appendix

\section{Symmetry group condition\label{sec:cos}}
%%%%%%%%%%%%%%%%%%%%%%%%%%%%%%%%%%%%%%%%%%%%%%%%%%%%%%%%%%%%%%%%%%%%%%%%

%%%%%%%%%%%%%%%%%%%%%%%%%%%%%%%%%%%%%%%%%%%%%%%%%%%%%%%%%%%%%%%%%%%%%%%%%%%%%%%%%
All the $T$'s and $R$'s in Eq.~(\ref{eq:gut-80}) and
Eq.~(\ref{eq:gut-81-1}) are $SU(3)$ matrices, {\it i.e.} $\det T=\det R=1$,
which implies that the product $T_{i}R_{j}$ is also an $SU(3)$
matrix.
One can use properties of $SU(3)$ matrices to derive the relations below.

Introducing the eigenvalues of $W_{ij} \equiv T_{i}R_{j}$
$(\lambda_{1},\thinspace\lambda_{2},\thinspace\lambda_{3})$ we can represent
this matrix as
\begin{equation}
W_{ij} = T_{i}R_{j}=U_{TS}\left(\begin{array}{ccc}
\lambda_1\\
 & \lambda_2\\
 &  & \lambda_3
\end{array}\right)U_{TS}^{\dagger}.
\label{eq:gut-84}
\end{equation}
Then according to (\ref{eq:gut-83}), $\lambda_i^p = 1$.
Keeping in mind that the eigenvalues satisfy the relations
$|\lambda_{1}|^{2}=|\lambda_{2}|^{2} =
|\lambda_{3}|^{2}=\lambda_{1}\lambda_{2}\lambda_{3}=1$ (the latter follows from
$\det W_{ij} = 1$), we can parametrize them as
%%the product $T_{i}R_{j}$ can be diagonalized by
\begin{equation}
\lambda_1 = e^{-i(\theta_{2}+\theta_{3})}, ~~
\lambda_2 =  e^{i\theta_{2}}, ~~
\lambda_3 = e^{i\theta_{3}}
\label{eq:eigen-par}
\end{equation}
with
\begin{equation}
\theta_i = \frac{2\pi n_i }{p}.
\label{eq:eigen-par1}
\end{equation}

%%Therefore Eq.~(\ref{eq:gut-83}) is equivalent to
%%\[
%%e^{ip\theta_{2}}=e^{ip\theta_{3}}=1.
%%\]

Since unitary transformations do not change the trace of a matrix,
%% The trace of $T_{i}R_{j}$,
we have% [according to Eq.~(\ref{eq:gut-84})]
\begin{equation}
{\rm tr}\left(T_{i}R_{j}\right)= \sum_i \lambda_i =
e^{-i(\theta_{2}+\theta_{3})}+e^{i\theta_{2}}+e^{i\theta_{3}}.
\label{eq:gut-85}
\end{equation}
For a given $p$ this sum has discrete sets of values.

On the other hand,
%%defining elements of matrix $U_0$,
%%\begin{equation}
%%U_{0}=\left(\begin{array}{ccc}
%%u_{11} & u_{12} & u_{13}\\
%%u_{21} & u_{22} & u_{23}\\
%%u_{31} & u_{32} & u_{33}
%%\end{array}\right),\label{eq:gut-82}
%%\end{equation}
we compute the trace  of  $T_{i}R_{j}$  from Eq.~(\ref{eq:gut-80}) and Eq.~(\ref{eq:gut-81-1}):
\begin{equation}
{\rm tr}\left(T_{i}R_{j}\right)=4(U_0)_{ij} (U_0)_{ij}^{*}-1   = 4 |(U_0)_{ij}|^2 - 1,
\label{eq:gut-86}
\end{equation}
which is a real number. Therefore, Eq.~(\ref{eq:gut-85}) must be
a real number too, which requires that
\begin{equation}
\sin\theta_{2}+\sin\theta_{3}-\sin(\theta_{2}+\theta_{3})=0.
\label{eq:gut-87}
\end{equation}
This equation has only three solutions in the range $[0,\thinspace2\pi)$:
\begin{equation}
 \theta_{2}=0,\ \ ~  \theta_{3} = 0, \ \  \theta_{2}+\theta_{3}=0,
\label{eq:gut-88}
\end{equation}
or equivalently, $n_1 = 0$, $n_2 = 0$,  $n_3 = 0$.
Consequently, in any of these three cases, the eigenvalues of $T_{i}R_{j}$
can be  taken as $(1,\ e^{2 \pi n/p },\ e^{- 2 \pi n/p} )$, and
thus the trace equals
\begin{equation}
{\rm tr}\left(T_{i}R_{j}\right) = 1 + 2\cos \frac{2 \pi n_{ij}}{p_{ij}}
= 4 \cos^2 \frac{\pi n_{ij}}{p_{ij}}-1.
\label{eq:gut-89}
\end{equation}
Using Eq.~(\ref{eq:gut-86}) and Eq.~(\ref{eq:gut-89}) we immediately obtain  equality (\ref{eq:gut-27}).

\section{The $S_{4}$ group and its representations\label{sec:S4}}

The group $S_{4}$ is the permutation group of four objects.
It can be defined by  four generators $r_{1}$, $r_{2}$, $t_{1}$
and $t_{2}$ with the following relations:
\begin{equation}
r_{1}^{2}=r_{2}^{2}=t_{1}^{2}=t_{2}^{2}=1,\label{eq:gut-110}
\end{equation}
\begin{equation}
(t_{1}r_{1})^{4}=(t_{1}r_{2})^{4}=(t_{2}r_{1})^{3}=(t_{2}r_{2})^{3}=1.\label{eq:gut-111}
\end{equation}
The group has five irreducible representations, denoted as $\mathbf{1}$,
$\mathbf{1}'$,$\mathbf{2}$, $\mathbf{3}$ and $\mathbf{3}'$, in
which the generators are represented by the following matrices:
\begin{equation}
R_{1}^{(\mathbf{3})}=-\frac{1}{2}\left(\begin{array}{ccc}
0 & \sqrt{2} & \sqrt{2}\\
\sqrt{2} & 1 & -1\\
\sqrt{2} & -1 & 1
\end{array}\right),\ R_{2}^{(\mathbf{3})}=-\left(\begin{array}{ccc}
1 & 0 & 0\\
0 & 0 & 1\\
0 & 1 & 0
\end{array}\right),\ T_{1}^{(\mathbf{3})}=\left(\begin{array}{ccc}
1 & 0 & 0\\
0 & -1 & 0\\
0 & 0 & -1
\end{array}\right),\ T_{2}^{(\mathbf{3})}=\left(\begin{array}{ccc}
-1 & 0 & 0\\
0 & 1 & 0\\
0 & 0 & -1
\end{array}\right);
\label{eq:gut-16}
\end{equation}
\begin{equation}
R_{1}^{(\mathbf{3}')}=\left(\begin{array}{ccc}
0 & 1 & 0\\
1 & 0 & 0\\
0 & 0 & 1
\end{array}\right),\ R_{2}^{(\mathbf{3}')}=\left(\begin{array}{ccc}
-1 & 0 & 0\\
0 & -1 & 0\\
0 & 0 & 1
\end{array}\right),\ T_{1}^{(\mathbf{3}')}=\left(\begin{array}{ccc}
1 & 0 & 0\\
0 & -1 & 0\\
0 & 0 & -1
\end{array}\right),\ T_{2}^{(\mathbf{3}')}=\left(\begin{array}{ccc}
1 & 0 & 0\\
0 & 0 & 1\\
0 & 1 & 0
\end{array}\right);
\label{eq:gut-17}
\end{equation}
\begin{equation}
R_{1}^{(\mathbf{2})}=\left(\begin{array}{cc}
-1 & 0\\
0 & 1
\end{array}\right),\ R_{2}^{(\mathbf{2})}=\left(\begin{array}{cc}
1 & 0\\
0 & 1
\end{array}\right),\ T_{1}^{(\mathbf{2})}=\left(\begin{array}{cc}
1 & 0\\
0 & 1
\end{array}\right),\ T_{2}^{(\mathbf{2})}=\left(\begin{array}{cc}
\frac{1}{2} & \frac{\sqrt{3}}{2}e^{-\frac{\pi i}{6}}\\
\frac{\sqrt{3}}{2}e^{\frac{\pi i}{6}} & -\frac{1}{2}
\end{array}\right),
\label{eq:gut-18}
\end{equation}
where $\omega\equiv\exp\left(\frac{2\pi i}{3}\right)$, and
\begin{equation}
R_{1}^{(\mathbf{1}')}=-1,\ R_{2}^{(\mathbf{1}')}=1,\ T_{1}^{(\mathbf{1}')} =
1,\ T_{2}^{(\mathbf{1}')}=-1;
\label{eq:gut-19}
\end{equation}
\begin{equation}
R_{1}^{(\mathbf{1})}=1,\ R_{2}^{(\mathbf{1})}=1,\ T_{1}^{(\mathbf{1})} = 1,
\ T_{2}^{(\mathbf{1})}=1.
\label{eq:gut-20}
\end{equation}

The Clebsch-Gordan (CG) coefficients are given by
\begin{eqnarray}
\left(\begin{array}{c}
x_{1}\\
x_{2}\\
x_{3}
\end{array}\right)^{(\mathbf{3})}\otimes\left(\begin{array}{c}
y_{1}\\
y_{2}\\
y_{3}
\end{array}\right)^{(\mathbf{3})} & = & \left(\begin{array}{c}
\frac{x_{1}y_{1}+x_{2}y_{2} +
x_{3}y_{3}}{\sqrt{3}}\end{array}\right)^{(\mathbf{1})}\oplus\frac{1}{6\sqrt{2}}\left(\begin{array}{c}
6x_{1}y_{1}-3\left(x_{2}+x_{3}\right)\left(y_{2}+y_{3}\right)\\
e^{\frac{\pi i}{6}}\sqrt{3}\left(2x_{1}y_{1} -
x_{2}\left(y_{2}-3y_{3}\right)+x_{3}\left(3y_{2}-y_{3}\right)\right)
\end{array}\right)^{(\mathbf{2})}\nonumber \\
 &  & \oplus\frac{1}{\sqrt{2}}\left(\begin{array}{c}
x_{3}y_{2}-x_{2}y_{3}\\
x_{1}y_{3}-x_{3}y_{1}\\
x_{2}y_{1}-x_{1}y_{2}
\end{array}\right)^{(\mathbf{3})}\oplus\frac{1}{2}\left(\begin{array}{c}
\sqrt{2}\left(x_{2}y_{2}-x_{3}y_{3}\right)\\
x_{2}y_{1}-x_{3}y_{1}+x_{1}\left(y_{2}-y_{3}\right)\\
-\left(x_{2}y_{1}+x_{3}y_{1}+x_{1}\left(y_{2}+y_{3}\right)\right)
\end{array}\right)^{(\mathbf{3}')},
\label{eq:gut-21}
\end{eqnarray}

\begin{eqnarray}
\left(\begin{array}{c}
x_{1}\\
x_{2}\\
x_{3}
\end{array}\right)^{(\mathbf{3}')}\otimes\left(\begin{array}{c}
y_{1}\\
y_{2}\\
y_{3}
\end{array}\right)^{(\mathbf{3}')} & = & \left(\begin{array}{c}
\frac{x_{1}y_{1}+x_{2}y_{2}+
x_{3}y_{3}}{\sqrt{3}}\end{array}\right)^{(\mathbf{1})}\oplus\frac{1}{\sqrt{2}}\left(\begin{array}{c}
x_{1}y_{1}-x_{2}y_{2}\\
\frac{e^{\frac{\pi i}{6}}\left(x_{1}y_{1}+x_{2}y_{2}-2x_{3}y_{3}\right)}{\sqrt{3}}
\end{array}\right)^{(\mathbf{2})}\nonumber \\
 &  & \oplus\frac{1}{2}\left(\begin{array}{c}
\sqrt{2}\left(-x_{3}y_{2}+x_{2}y_{3}\right)\\
\left(x_{2}y_{1}+x_{3}y_{1}-x_{1}\left(y_{2}+y_{3}\right)\right)\\
\left(-x_{2}y_{1}+x_{3}y_{1}+x_{1}\left(y_{2}-y_{3}\right)\right)
\end{array}\right)^{(\mathbf{3})}\oplus\frac{1}{\sqrt{2}}\left(\begin{array}{c}
\left(x_{3}y_{2}+x_{2}y_{3}\right)\\
\left(x_{3}y_{1}+x_{1}y_{3}\right)\\
\left(x_{2}y_{1}+x_{1}y_{2}\right)
\end{array}\right)^{(\mathbf{3}')},
\label{eq:gut-21-1}
\end{eqnarray}
\begin{equation}
\left(\begin{array}{c}
x_{1}\\
x_{2}
\end{array}\right)^{(\mathbf{2})}\otimes\left(\begin{array}{c}
y_{1}\\
y_{2}
\end{array}\right)^{(\mathbf{2})} =
\left(\frac{x_{1}y_{1}-\omega x_{2}y_{2}}{\sqrt{2}}\right)^{(\mathbf{1})}\oplus
\left(\frac{x_{1}y_{2} -
x_{2}y_{1}}{\sqrt{2}}\right)^{(\mathbf{1}')}\oplus\frac{1}{\sqrt{2}}
\left(\begin{array}{c}
x_{2}y_{1}+x_{1}y_{2}\\
-\omega^{2}x_{1}y_{1}-x_{2}y_{2}
\end{array}\right)^{(\mathbf{2}')}.\label{eq:gut-22}
\end{equation}

\section{Analytic diagonalization in the visible sector\label{sec:analytic}}
%%%%%%%%%%%%%%%%%%%%%%%%%%%%%%%%%%%%%%%%%%%%%%%%%%%%%%%%%%%%%%%%%%%%%%%%%%%%%%%

%By plugging $V_{d}$ and $U_{e}$ back into

To diagonalize the mass matrices in Eq.~(\ref{eq:gut-41}), we reconstruct the mass matrices of down quarks and charged leptons in
terms of mass eigenstates and mixing angles:
$M_{d}=U_{d}{\rm diag}(\tilde{m}_{d}, \thinspace \tilde{m}_{s},
\thinspace \tilde{m}_{b})U_{d}^{T}$
and $M_{l}=
U_{l}{\rm diag}(\tilde{m}_{e},\thinspace \tilde{m}_{\mu},\thinspace
\tilde{m}_{\tau})U_{l}^{T}$, which gives
%
%and equating the matrices in terms of mass parameters $d_i$, $k_i$
%  to we get
%xxj: f->f e^{i\phi_f}
\begin{equation}
\left(\begin{array}{cc}
\tilde{m}_{d}c^{2}+\tilde{s}^{2}\tilde{m}_{s} &
c(\tilde{s}\tilde{m}_{s}-\tilde{m}_{d}\tilde{s}^{*})\\
c(\tilde{s}\tilde{m}_{s}-\tilde{m}_{d}\tilde{s}^{*}) &
\tilde{m}_{s}c^{2}+\tilde{m}_{d}\tilde{s}^{*2}
\end{array}\right)=\left(\begin{array}{cc}
d_{1}+k_{1} & f e^{i\phi_f}\\
f e^{i\phi_f}  & d_{2}+k_{2}
\end{array}\right),
\label{eq:gut-56}
\end{equation}
\begin{equation}
\left(\begin{array}{cc}
\tilde{m}_{e}c_{e}^{2}+\tilde{s}_{e}^{2}\tilde{m}_{\mu} &
c_{e}(\tilde{s}_{e}\tilde{m}_{\mu}-\tilde{m}_{e}\tilde{s}_{e}^{*})\\
c_{e}(\tilde{s}_{e}\tilde{m}_{\mu}-
\tilde{m}_{e}\tilde{s}_{e}^{*}) & \tilde{m}_{\mu}c_{e}^{2} +
\tilde{m}_{e}\tilde{s}_{e}^{*2}
\end{array}\right)=\left(\begin{array}{cc}
-3d_{1}+k_{1} & -3f e^{i\phi_f} \\
-3f e^{i\phi_f} & -3d_{2}+k_{2}
\end{array}\right).
\label{eq:gut-57}
\end{equation}
%%Here the $\tilde{m}$'s are masses attached with complex phases
%%(e.g. $\tilde{m}_{s}=m_{s}e^{i\alpha_{s}}$).
Because $\tilde{m}_{d}\tilde{s}^{*2}\ll\tilde{m}_{s}c^{2}$
and $\tilde{m}_{e}\tilde{s}_{e}^{*2}\ll\tilde{m}_{\mu}c_{e}^{2}$,
we neglect $\tilde{m}_{d}\tilde{s}^{*2}$ and $\tilde{m}_{e}\tilde{s}_{e}^{*2}$
below.
%%We first focus on the diagonal elements.
Equating the corresponding elements of the matrices on the left-hand side and right-hand side
we obtain expressions for $d_{1,2}$ and $k_{1,2}$:
\begin{equation}
k_{1}=\frac{1}{4}\left(3c^{2}\tilde{m}_{d} +
c_{l}^{2}\tilde{m}_{e}+\tilde{m}_{\mu}\tilde{s}_{l}^{2} +
3\tilde{m}_{s}\tilde{s}^{2}\right),\ \
k_{2} =
\frac{1}{4}(3c^{2}\tilde{m}_{s} + c_{l}^{2}\tilde{m}_{\mu}),
\label{eq:gut-58}
\end{equation}
\begin{equation}
d_{1}=\frac{1}{4}\left(c^{2}\tilde{m}_{d} -
c_{l}^{2}\tilde{m}_{e}-\tilde{m}_{\mu}\tilde{s}_{l}^{2}+
\tilde{m}_{s}\tilde{s}^{2}\right),\ \ d_{2}=\frac{1}{4}(c^{2}\tilde{m}_{s}-
c_{l}^{2}\tilde{m}_{\mu}).
\label{eq:gut-59}
\end{equation}
Since $-3d_{2}+k_{2}$ and $d_{2}+k_{2}$ in Eq.~(\ref{eq:gut-57})
and Eq.~(\ref{eq:gut-56}) are real under our assumptions,
$\tilde{m}_{s}$ and $\tilde{m}_{\mu}$
should be approximately real, either positive or negative. Because
$d_{2}$ is dominant, we take positive $\tilde{m}_{s}$ and negative
$\tilde{m}_{\mu}$,  {\it i.e.} $\tilde{m}_{s}\approx m_{s}$
and $\tilde{m}_{\mu}\approx-m_{\mu}$.
Furthermore, the equalities
${\rm Im}(d_{1}+k_{1}) = {\rm Im}(-3d_{1}+k_{1}) = 0$
give
\begin{equation}
c^{2}m_{d}\sin\alpha_{d}+s^{2}m_{s}\sin2\phi_{C} =
c_{e}^{2}m_{e}\sin\alpha_{e} -s_{l}^{2}m_{\mu}\sin2\phi_{l}=0,
\label{eq:gut-60}
\end{equation}
%%where $\alpha_{d}$ and $\alpha_{e}$ are the phases in $\tilde{m}_{d}$
%%and $\tilde{m}_{e}$ respectively. From Eq.~(\ref{eq:gut-60}), they
%%are determined by
which lead to
\begin{equation}
\sin\alpha_{d}=-\frac{s^{2}m_{s}}{c^{2}m_{d}}\sin2\phi_{C},\ \ \
\sin\alpha_{e}=\frac{s_{l}^{2}m_{\mu}}{c_{l}^{2}m_{e}}\sin2\phi_{l}.\
\label{eq:gut-62}
\end{equation}

From the equality of the off-diagonal elements of (\ref{eq:gut-56})
(\ref{eq:gut-57}), we obtain
\begin{equation}
f e^{i\phi_f} =c(\tilde{s}m_{s}-\tilde{m}_{d}\tilde{s}^{*}) =
\frac{1}{3}c_{l}(\tilde{s}_{l}m_{\mu}+\tilde{m}_{e}\tilde{s}_{e}^{*}),
\label{eq:gut-63}
\end{equation}
or
\begin{equation}
2f e^{i\phi_f} = \sin2\theta_{C}(m_{s}e^{i\phi_{C}} -
\tilde{m}_{d}e^{-i\phi_{C}})=\frac{1}{3}\sin2\theta_{e}(m_{\mu}e^{i\phi_{l}} +
\tilde{m}_{e}e^{-i\phi_{l}}).
\label{eq:gut-64}
\end{equation}
Because $m_{s}\gg m_{d}$ and $m_{\mu}\gg m_{e}$,
one can immediately see from (\ref{eq:gut-64}) that the phases $\phi_{C}$
and $\phi_{l}$ should be approximately equal to $\phi_{f}$:
\begin{equation}
\phi_{C} = \phi_{f}+{\cal O}\left(\frac{m_{d}}{m_{s}}\right),\ \
\phi_{l}=\phi_{f}+{\cal O}\left(\frac{m_{e}}{m_{\mu}}\right).\
\label{eq:gut-61}
\end{equation}
This reproduces the result in Eq.~(\ref{eq:gut-65}). Eq.~(\ref{eq:gut-64})
also gives
\begin{equation}
\frac{\sin2\theta_{l}}{\sin2\theta_{C}} =
\frac{3(m_{s}e^{i\phi_{C}}-\tilde{m}_{d}e^{-i\phi_{C}})}{m_{\mu}e^{i\phi_{e}} +
\tilde{m}_{e}e^{-i\phi_{l}}}\approx
\frac{3(m_{s} + m_{d}e^{i\phi_{1}})}{m_{\mu} + m_{e}e^{i\phi_{2}}},
\label{eq:gut-67}
\end{equation}
\begin{equation}
\phi_{1} \equiv \pi+\alpha_{d}-2\phi_{f},\
\phi_{2}\equiv\alpha_{e}-2\phi_{f},
\label{eq:gut-55}
\end{equation}
where we have taken the approximation $\phi_{C}\approx\phi_{f}\approx\phi_{e}$.
 %and introduced the definition

Due to the relations  $m_{s}\gg m_{d}$ and $m_{\mu}\gg m_{e}$, in Eq.~(\ref{eq:gut-67})
the imaginary parts in $m_{d}e^{i\phi_{1}}$ and $m_{e}e^{i\phi_{2}}$
can be neglected, which leads to
Eq.~(\ref{eq:gut-66}).

Finally, we express $d_{1,2}$, $k_{1,2}$
and $f$ in terms of
$(m_{d},\thinspace m_{s},\thinspace m_{e},\thinspace m_{\mu},
\thinspace\theta_{C},\thinspace\phi_{l})$ and $\theta_l$
%%which are more straightforward observables. By plugging
inserting results of Eqs.~(\ref{eq:gut-61}),
and (\ref{eq:gut-62}) into Eqs.~(\ref{eq:gut-58}), (\ref{eq:gut-59})
and (\ref{eq:gut-63}):
\begin{equation}
k_{1}=\frac{1}{4}\left(3\sqrt{c^{4}m_{d}^{2}- s^{4}m_{s}^{2}\sin^{2}2\phi_{l}} +
\sqrt{c_{l}^{4}m_{e}^{2} - s_{l}^{4}m_{\mu}^{2}\sin^{2}
\left(2\phi_{l}\right)} + \cos 2\phi_{l}
\left(3s^{2}m_{s}-s_{e}^{2}m_{\mu}\right)\right),
\label{eq:gut-68}
\end{equation}
\begin{equation}
d_{1}=\frac{1}{4}\left(\sqrt{c^{4}m_{d}^{2} -
s^{4}m_{s}^{2}\sin^{2} 2\phi_{l}}-\sqrt{c_{l}^{4}m_{e}^{2}- s_{l}^{4}m_{\mu}^{2}
\sin^{2} 2\phi_{l}} + \cos 2\phi_{l} \left(s_{l}^{2}m_{\mu} + s^{2}m_{s}\right)\right),
\label{eq:gut-69}
\end{equation}
\begin{equation}
k_{2}=\frac{1}{4}\left(3c^{2}m_{s}-c_{l}^{2}m_{\mu}\right),
\label{eq:gut-70}
\end{equation}
\begin{equation}
d_{2}=\frac{1}{4}\left(c^{2}m_{s} + c_{l}^{2}m_{\mu}\right),
\label{eq:gut-71}
\end{equation}
\begin{equation}
f e^{i\phi_f}=sc
\left[m_{s}e^{i\phi_{C}} -
m_{d} \exp \left(-i \alpha_d - i\phi_{C} \right) \right].
\label{eq:gut-72}
\end{equation}
Here
\begin{equation}
\alpha_d = - \arcsin \left(\frac{s^{2}m_{s}}{c^{2}m_{d}}\sin2\phi_{C}\right).
\end{equation}

%%The above expressions have expressed $d_{1,2}$, $k_{1,2}$ and $f$
%%in terms of $(m_{d},\thinspace m_{s},\thinspace m_{e},\thinspace m_{\mu},
%%\thinspace\theta_{C},\thinspace\phi_{e})$ and $\theta_{e}$.
The lepton mixing can be obtained from (\ref{eq:gut-67})
\begin{equation}
\sin2\theta_{l}\approx3\sin2\theta_{C}\left|\frac{m_{s} -
e^{-2i\phi_{l}}e^{-i\sin^{-1}\left(\frac{s^{2}\sin\left(2\phi_{l}\right)
m_{s}}{c^{2}m_{d}}\right)}m_{d}}{m_{\mu} +
e^{-2i\phi_{l}}e^{i\sin^{-1}\left(\frac{\sin\left(2\phi_{l}\right)
m_{\mu}s_{l}^{2}}{c_{e}^{2}m_{e}}\right)}m_{e}}\right|,
\label{eq:gut-73}
\end{equation}
where the right-hand side still contains $\theta_{l}$ but it only
appears in the negligibly small term proportional to $m_{e}$. So one can simply replace $\theta_{l}$ by $\theta_{C}$ in this term.

%To numerically check the accuracy of the above approximate equations,
%we take
%\begin{equation}
%(m_{d},\thinspace m_{s},\thinspace m_{e},\thinspace m_{\mu}) =
%(1.24, \thinspace21.7,\thinspace0.3585,\thinspace75.67){\rm MeV},\
%s=0.225,\ \phi_{l}=-90^{\circ},
%\label{eq:gut-74}
%\end{equation}
%to evaluate $d_{1,2},k_{1,2}$ and $f$:
%\begin{equation}
%(k_{1},\thinspace k_{2},\thinspace d_{1}, \thinspace d_{2},
%\thinspace f e^{i\phi_f})=(0.9373,\thinspace-2.674,\thinspace -
%0.8586,\thinspace23.28,\thinspace3.35\times10^{-16}+5.029i){\rm MeV}.
%\label{eq:gut-75}
%\end{equation}
%Using the values in Eq.~(\ref{eq:gut-75}) to reconstruct $M_{d}$
%and $M_{e}$ and then diagonalizing them according to %Eq.~(\ref{eq:gut-35}) and
%Eq.~(\ref{eq:gut-36}), one can obtain
%\[
%(m_{d},\thinspace m_{s},\thinspace m_{e},\thinspace m_{\mu}) =
%(1.237,\thinspace21.76,\thinspace0.3579,\thinspace75.65){\rm MeV},\ s=0.224,\
%\phi_{l}=-90.00^{\circ},
%\]
%which agrees with Eq.~(\ref{eq:gut-74}) very well.
%Therefore our
We have checked that our
analytic results agree with numerical computations up to an order of $10^{-4}$.
%should be accurate up to the order of $10^{-4}$.

\bibliographystyle{apsrev4-1}
\bibliography{ref}

\end{document}